\documentclass[aps,twocolumn,showpacs,superscriptaddress,preprintnumbers,amsmath,amssymb,longbibliography]{revtex4-1}

\pdfoutput=1

\usepackage{graphicx}
\usepackage{dcolumn}
\usepackage{bm}
\usepackage{color} 
\usepackage{subfigure} 	
\usepackage{hyperref}
\usepackage{ulem}

\begin{document}


\title{Tunable superconducting two-chip lumped element resonator}

\author{B.~Ferdinand}
\email{benedikt-martin.ferdinand@uni-tuebingen.de}
\author{D.~Bothner}
\thanks{present adress: Kavli Institute of Nanoscience, Delft University of Technology, PO Box 5046, 2600 GA, Delft, The Netherlands}
\author{R.~Kleiner}
\author{D.~Koelle}%
\affiliation{%
Physikalisches Institut and Center for Quantum Science (CQ) in LISA$^{+}$,
Universit\"{a}t
T\"{u}bingen, Auf der Morgenstelle 14, D-72076 T\"{u}bingen, Germany
}%
\date{\today}

\begin{abstract}

We have fabricated and investigated a stacked two-chip device, consisting of a lumped element resonator on one chip, which is side-coupled to a coplanar waveguide transmission line on a second chip. We present a full model to predict the behavior of the device dependent on the position of the lumped element resonator with respect to the transmission line. We identify different regimes, in which the device can be operated. One of them can be used to tune the coupling between the two subsystems. Another regime enables frequency tunability of the device, without leaving the over-coupled limit for internal quality factors of about $10^4$, while in the last regime the resonator properties are insensitive against small variations of the position. Finally, we have measured the transmission characteristics of the resonator for different positions, demonstrating a good agreement with the model.

\end{abstract}

\maketitle
\section{Introduction}
Superconducting coplanar microwave resonators have gained increasing interest during the past years. In particular, they represent a main building block of the architecture of circuit quantum electrodynamics (cQED) \cite{blais2004cavity,wallraff2004strong,wallraff2005,fink2008climbing,hofheinz2009synthesizing,niemczyk2010circuit,wilson2011observation,kelly2015state,bosman2017multi}. Despite the breakthroughs in cQED, these strongly coupled systems still suffer from short coherence times. Thus, hybrid quantum systems have been proposed, which combine superconducting qubits and natural spin-systems \cite{andre2006coherent,rabl2006hybrid,hafezi2012atomic,petrosyan2008quantum,verdu2009strong,imamouglu2009cavity,henschel2010cavity}. Such systems have been reported for ensembles of electron spins in diamond \cite{schuster2010high,kubo2010strong,wu2010storage,amsuss2011cavity}, erbium \cite{probst2013anisotropic} or phosphorus donors \cite{zollitsch2015high}. For ultracold atomic clouds, coherence times of several seconds have been demonstrated close to a superconducting resonator \cite{bernon2013manipulation} and the coupling between such a cloud and a superconducting resonator has been realized \cite{Hattermann2017Coupling}. 

The resonators used in most cQED experiments can basically be separated in two types of resonators: coplanar waveguide (CPW) resonators and lumped element (LE) resonators. The latter can be used to increase the magnetic (electric) coupling to spin-systems, by designing them to have a low (high) impedance \cite{samkharadze2016high,bienfait2016controlling,sarabi2017, stockklauser2017strong,bosman2017approaching}. Some of these spin-systems, e.g. cold atoms in a magnetic trap, have a nearly fixed energy spectrum, which demands for a tunable resonator in order to study both the resonant and the non-resonant interaction between the resonator and the spin-system. The LE resonators are typically side-coupled to a transmission line (TL) for excitation and readout \cite{geerlings2012,deng2013}. However, both the TL and the LE resonator are patterned on the same substrate, which requires the additional implementation of e.g. tunable inductors \cite{palacios2008tunable,sandberg2008tuning,vissers2015frequency} in order to make the device tunable with respect to both its resonance frequency and its coupling to the input-output-circuit. Tunable inductors are typically realized using superconducting quantum interference devices (SQUIDs) or SQUID arrays \cite{palacios2008tunable,sandberg2008tuning}. However, in, e.g., electron paramagnetic resonance experiments or for hybrid systems consisting of a superconducting microwave circuitry and ultracold atomic ensembles, the operation conditions make the use of SQUIDs unpractical for two reasons. First, numerous flux quanta would be trapped in the SQUID ring and the critical currents of the SQUID junctions would presumably be strongly reduced under the influence of out-of-plane magnetic fields in the mT-range. Second, at high photon numbers such resonators would show a considerable nonlinear behavior due to their low dynamic range.


In this work, we demonstrate a tunable stacked two-chip device with a LE resonator with a linear inductor on one chip, coupled to a CPW TL with finite width ground planes on a second chip. Tuning both the resonance frequency and the coupling between the TL and the resonator can be realized by simply moving the LE resonator with respect to the TL. From the technical side, the two-chip approach allows for fast exchange of the resonator chips, since they are galvanically decoupled from the TL and thus do not require additional electrical contact. A full model is employed by virtue of which the resonator properties can be predicted, enabling precise designing opportunities. We measure the transmission response of a resonator with a resonance frequency of about $5.86\,\textrm{GHz}$ and compare the results with the prediction of the model as the position of the resonator chip is varied. We find that model and measurement agree very well. The analysis of the resonator properties as a function of the position reveals different regimes in which the device can be operated. In particular, in one regime, the resonance frequency remains almost constant, whereas the coupling between resonator and TL is varied significantly, and vice versa. Thereby, the resonance frequency can be tuned by $25\,\textrm{MHz}$ without causing significant shifts of the internal quality factor and the coupling between resonator and TL can be switched on and off, without shifting the resonance frequency strongly. Thus, the total decay rate can be controlled, which is crucial for pushing, e.g., a hybrid quantum system with ultracold atomic gases towards the strong coupling regime, where relatively weak coupling strengths of about $20-50\,\textrm{kHz}$ are expected. Accordingly, the presented tunable device can be a useful tool for experiments where the frequency is adjusted prior to the experiment and hence, tuning speed is not a relevant parameter.

\section{Fabrication and sample design}
\label{sec:fab}
\begin{figure}[hb]
			\includegraphics[width=0.45\textwidth]{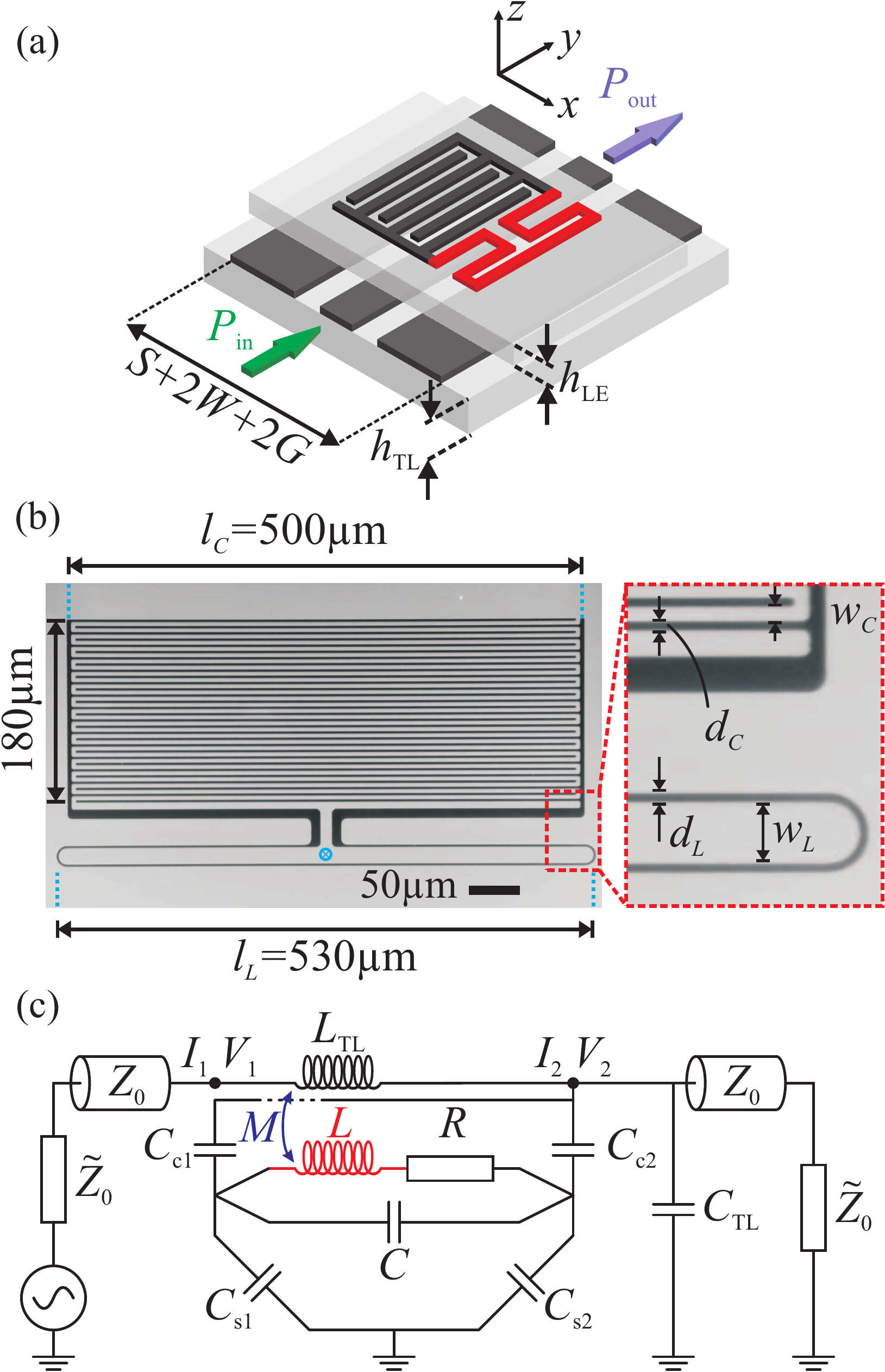}
	\caption{Device and equivalent circuit. (a) Schematic view of the experimental setup, with the LE-chip mounted on top of the TL-chip. Dark grey parts on the top chip indicate the capacitor, red parts show the inductive loop of the circuit. (b) Optical image of a LE resonator (here: $l_C=500\,\mathrm{\mu m}$, $l_L=530\,\mathrm{\mu m}$), the blue circle indicates the center of the loop. The zoomed version shows a section of the cigar shaped inductor and of the interdigital capacitor. (c) Circuit representation of the lumped device (for details see text).}
	\label{fig:Scheme}
\end{figure}

Our device consists of two parts, the TL for excitation and readout, and the LE resonator. Each of these two elements is fabricated on a separate sapphire substrate. For the experiment, the LE-chip is mounted on top of the TL-chip (see Fig.~\ref{fig:Scheme}(a)). The fabrication processes of both chips in general follow the same routine, starting with the deposition of a $500\,\textrm{nm}$ thick layer of superconducting niobium (Nb, transition temperature $T_{c,\textrm{Nb}}=9.2\,\textrm{K}$), deposited by magnetron sputtering on a r-cut sapphire substrate with a thickness $h_\textrm{TL}=330\,\mathrm{\mu m}$ for the TL, and $h_\textrm{LE}=100\,\mathrm{\mu m}$ for the LE resonator (see Fig.~\ref{fig:Scheme}). For the resonator chip, we use a double-sided polished substrate to facilitate alignment and a visual position determination during the chip stacking process. After an optical lithography step, the structures are etched by means of reactive ion etching with $\mathrm{SF_6}$.   
A schematic of the system is shown in Fig.~\ref{fig:Scheme}(a). The TL has a typical CPW geometry with finite width ground planes. More precisely, the TL has a center conductor width of $S=200\,\mathrm{\mu m}$, a gap between the center conductor and the ground planes of $W=90\,\mathrm{\mu m}$ and a width of the ground planes of $G=400\,\mathrm{\mu m}$. The LE resonators consist of an interdigital capacitor (IDC, dark grey in Fig.~\ref{fig:Scheme}(a)) in parallel with an inductive loop (red in Fig.~\ref{fig:Scheme}(a)). The capacitor has $N=30$ fingers with a length of $l_C=500\,\mathrm{\mu m}$. The width of the fingers is $d_C=2.4\,\mathrm{\mu m}$ and the gap between neighboring fingers is $w_C=3.6\,\mathrm{\mu m}$. The inductive loop has two parts: first, the cigar shaped main part with a length of $l_L=530\,\mathrm{\mu m}$, a loop width $w_L=15\,\mathrm{\mu m}$ and a conductor width of $d_L=2.4\,\mathrm{\mu m}$, and second, the connection to the IDC. For an adequate description, however, the inductance of the capacitor fingers has also to be taken into account. An optical image of the resonator part can be seen in Fig.~\ref{fig:Scheme}(b).

Due to the double-sided polished substrate, the LE-chip can be moved in $x$-direction without scratching the Nb thin film of the TL (for details of the sample holder see Supplemental Material \cite{ferdinand2017SI}). Below we show that moving the resonator can be used to tune both the resonance frequency and the coupling to the TL. In particular, one can place the LE-chip at a "zero coupling position", i.e., at a position where no coupling to the TL can be observed. 

\section{Simulation and Model}
\label{sec:model}
Figure \ref{fig:Scheme} (c) shows the LE circuit, which we use to model our device. The circuit consists of the LE resonator, represented by the inductor $L$ and the capacitor $C$. The part of the TL along the resonator is modeled by the inductor $L_\textrm{TL}$ and the capacitor $C_\textrm{TL}$, which can be understood as the inductance (capacitance) per unit length $L_l$ ($C_l$) integrated along the LE resonator, such that $\sqrt{L_\textrm{TL}/C_\textrm{TL}}$ is the corresponding characteristic TL impedance. For regions outside the overlap with the LE resonator, but still below the LE substrate, the characteristic impedance of the TL is given by $Z_0=41\,\Omega$. The remaining parts of the TL and the connection to a network analyzer have an impedance of $\tilde{Z}_0=50\,\Omega$. Several capacitors are used in the model to account for the frequency shifts, induced by the presence of the TL ($C_\textrm{c1}$, $C_\textrm{c2}$ - center conductor of the TL to resonator; $C_\textrm{s1}$, $C_\textrm{s2}$ - resonator to ground). Due to the symmetry of our sample, we assume $C_\textrm{c1}=C_\textrm{c2}=:C_\textrm{c}$ , $C_\textrm{s1}=C_\textrm{s2}=:C_\textrm{s}$ in the following, resulting in a vanishing capacitive coupling of the resonator, due to the vanishing phase difference at both sides of the capacitor and inductor. The coupling to the TL, which is thus purely inductive, is mediated via a mutual inductance $M$. At a temperature of $T=4.2\,\textrm{K}$, resistive losses are dominant, which is taken into account via a resistance $R$ in series with the inductor $L$.
\begin{figure}[ht]
			\includegraphics[width=0.47\textwidth]{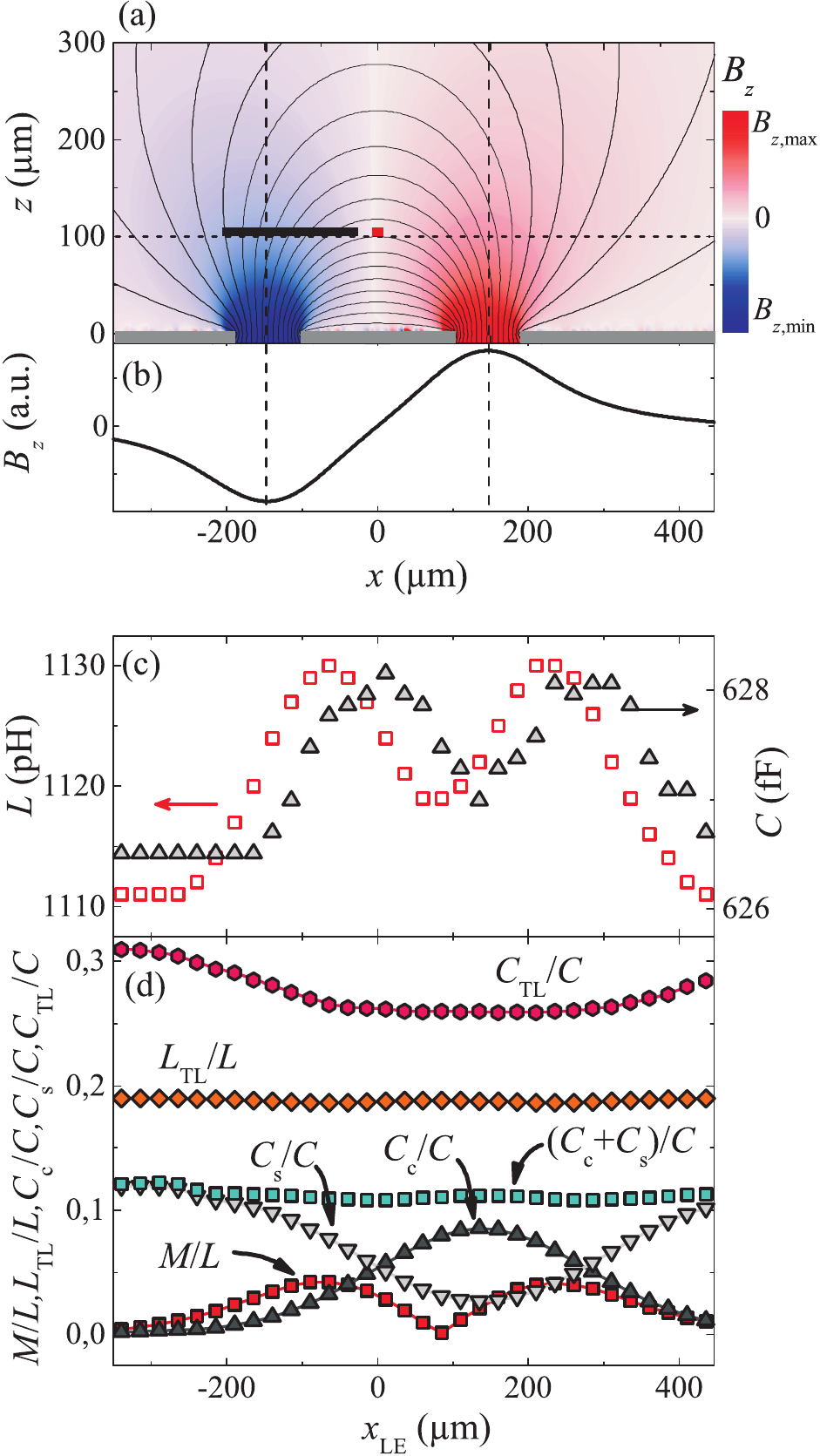}
	\caption{Simulation results of the circuit parameters. (a) Cross-sectional view of the magnetic field lines and $B_z$-component of the TL microwave magnetic field. Grey boxes picture the CPW, black and red box correspond to the width of the IDC and the inductive loop, respectively, for $x_\textrm{LE}=0$. Dashed line at $z=100\,\mathrm{\mu m}$ indicates plane for (b), linescan of $B_z$-component, (c), simulated variation of the inductance $L$ and the capacitance $C$, and (d), mutual inductance $M$, TL inductance $L_\textrm{TL}$, both normalized to $L$, as well as the capacitances $C_\textrm{c}$, $C_\textrm{s}$ and $C_\textrm{TL}$, normalized to $C$, for the relative distance $x_\textrm{LE}$ between the LE resonator and the TL. (for details see text)}
	\label{fig:Simu_L}
\end{figure}
The parameters of the device depend on the position of the LE resonator with respect to the TL. 

For the device presented in Sec.~\ref{sec:fab}, we perform static field simulations of these circuit parameters, as the relative position $x_\textrm{LE}$ of the LE resonator and the TL is varied in $x$-direction. The inductive parts were simulated using the numerical software package 3D-MLSI \cite{khapaev2002current}, which is based on the London and the Maxwell equations. For the simulation of the capacitance matrix we use COMSOL Multiphysics. The results of the simulations are shown in Fig.~\ref{fig:Simu_L}. Here, $x_\textrm{LE}=0$ corresponds to the position where the center of the inductor loop (blue circle in Fig.~\ref{fig:Scheme}(b)) is centered above the center conductor of the TL. Thereby, the capacitor is on the left side of the inductor (cf. black and red box in Fig.~\ref{fig:Simu_L}(a)). Varying the relative distance $x_\textrm{LE}$, the inductance $L$ is modified (see Fig.~\ref{fig:Simu_L}(c)). For positions $x_\textrm{LE}$ at which the inductive parts of the LE resonator are above a superconducting surface, the inductance $L$ decreases due to screening currents in the superconducting parts of the TL. These currents correspond to a mutual inductance which is mapped on $L$ and, thereby, reduces $L$. Note, that this mutual inductance is not the same as $M$, since the origin of this mutual inductance is the bare presence of a superconducting surface below the resonator, and the origin of the mutual inductance $M$ is the coupling to the TL. The magnetic coupling between TL and resonator is mainly mediated via the $z$-component of the magnetic field, or more precisely, by the flux threading the resonator. Figures \ref{fig:Simu_L}(a), (b) show that the $z$-component $B_z$ of the microwave magnetic field of the TL, and hence, the inductive coupling is expected to be strong directly above the gaps between the signal line and the ground conductor of the TL. Intuitively, one might assume that both the inductance $L$ and the mutual inductance $M$ are symmetric around $x_\textrm{LE}=0$. However, the symmetry axis is shifted towards $x_\textrm{LE}>0$, which can be seen in Figs.~\ref{fig:Simu_L}(c), (d). This is due to the fact, that the inductance of the IDC is taken into account in the simulations. The TL inductance $L_\textrm{TL}$ remains constant (Fig.~\ref{fig:Simu_L}(d)). Note, that Fig.~\ref{fig:Simu_L}(d) shows the position dependence of $M$ and $L_\textrm{TL}$, both normalized to $L$, as well as the capacitances $C_\textrm{c}$, $C_\textrm{s}$ and $C_\textrm{TL}$, all normalized to $C$. Accordingly, changes of the plotted normalized parameters with $x_\textrm{LE}$ may also reflect changes of $L$ and $C$ upon varying $x_\textrm{LE}$. However, Fig.~\ref{fig:Simu_L}(c) shows that the relative variation of $L$ and $C$ with $x_\textrm{LE}$ is $<2\%$ and $<0.3\%$, respectively. Hence, the shown dependencies in Fig.~\ref{fig:Simu_L}(d) reflect changes of $M$, $L_\textrm{TL}$, $C_\textrm{c}$, $C_\textrm{s}$ and $C_\textrm{TL}$ with $x_\textrm{LE}$.

In addition to the inductive changes, also the capacitance is modified for different positions $x_\textrm{LE}$. Whenever the IDC is above a conducting surface, part of the electric field lines end on this surface, reducing the IDCs capacitance (Fig.~\ref{fig:Simu_L}(c)). This is the case when the center of the loop is above the center of the gaps of the TL at $x_\textrm{LE}<-150\,\mathrm{\mu m}$ and $x_\textrm{LE}\approx150\,\mathrm{\mu m}$, and when the LE resonator is completely above the right ground conductor at $x_\textrm{LE}\approx400\,\mathrm{\mu m}$ (note, that the IDC is on the left side of the inductor). Furthermore, changing the position $x_\textrm{LE}$ of the resonator, the TL capacitance $C_\textrm{TL}$ is modified. For the same reason as for the capacitance $C$, the presence of other conducting parts, at positions of non-vanishing electric fields, leads to a reduction of the capacitance $C_\textrm{TL}$ (Fig.~\ref{fig:Simu_L}(d), $-100<x_\textrm{LE}<300\,\mathrm{\mu m}$). In Fig.~\ref{fig:Simu_L}(d) the variation of the capacitances $C_\textrm{c}$, $C_\textrm{s}$ is shown. The trend of $C_\textrm{c}$ along the position $x_\textrm{LE}$ is inverted to the one of $C_\textrm{s}$. Thus, the sum of these capacitances remains almost constant (cf. Fig.~\ref{fig:Simu_L}). As discussed in the Supplemental Material \cite{ferdinand2017SI} in more detail, this will result in a resonance frequency which has a weak dependence on the changes of the capacitances $C_\textrm{c}$ and $C_\textrm{s}$ of the circuit.

\begin{figure}[htb]
			\includegraphics[width=0.44\textwidth]{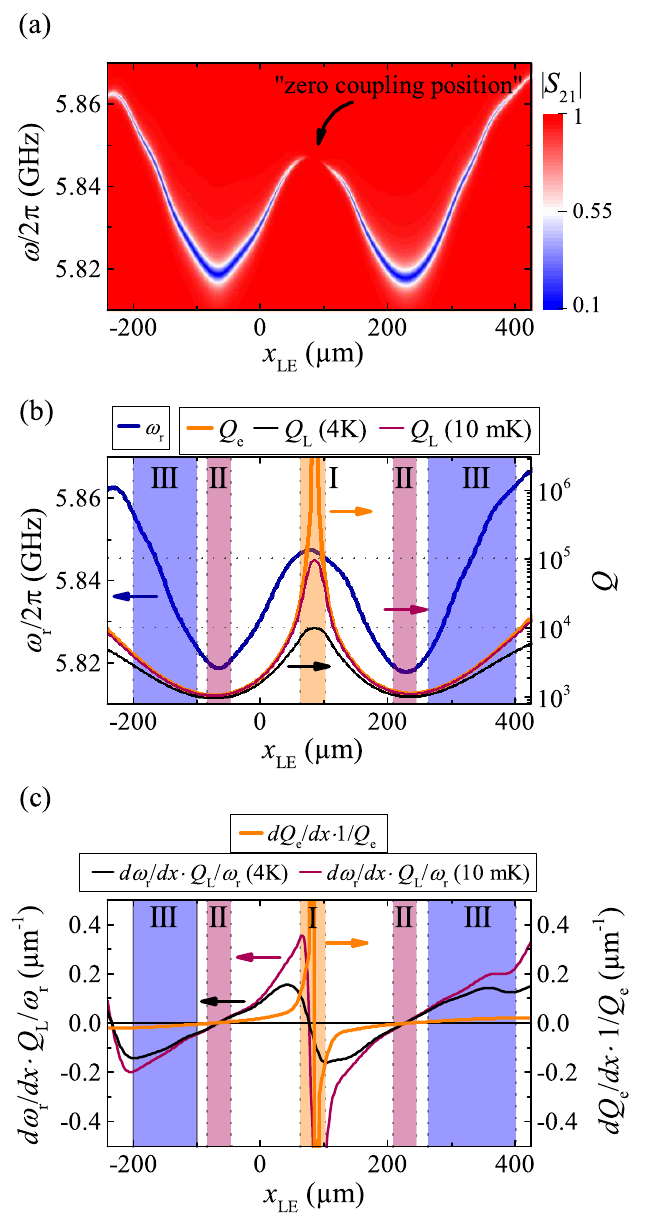}
	\caption{Simulated transmission behavior of the device and regimes for varying position $x_\textrm{LE}$. (a) Transmission amplitude $|S_{21}(\omega,x_\textrm{LE})|$ for $R=0.005\,\mathrm{\Omega}$ resulting in $Q_\textrm{i}=8200$, (b) resonance frequency (blue line; left axis) and external quality factor (orange line; right axis). In addition, we plot the loaded quality factor $Q_\textrm{L}$ for two different $Q_\textrm{i}$'s, taking realistic values for $4\,\textrm{K}$ (black line) and $10\,\textrm{mK}$ (red line). We assume here $Q_\textrm{i,4\,K}=10^4$ and $Q_\textrm{i,mK}=10^5$, indicated by dashed grey lines. (c) spatial derivative of the resonance frequency normalized to the cavity linewidth $\omega_\textrm{r}/Q_\textrm{L}$ (black line, $T=4\,\textrm{K}$; red line $T\approx10\,\textrm{mK}$), and relative change of the external quality factor. Colored areas correspond to regimes $\textbf{I}$ to $\textbf{III}$ (for details see text).}
	\label{fig:Model_1}
\end{figure}

Using Kirchhoff equations, we calculate the $ABCD$ matrix \cite{Poz98} of our circuit (cf. Fig.~\ref{fig:Scheme}(c)) and therewith the transmission function $S_{21}(\omega)$. The resonance frequency is found to obey
\begin{equation}
\omega_\textrm{r}=\frac{1}{\sqrt{LC}}\cdot\frac{1}{\sqrt{1+\gamma/2}}\,\, ,
\end{equation}
where $\gamma=(C_\textrm{c}+C_\textrm{s})\cdot C^{-1}$ (for details see Supplemental Material \cite{ferdinand2017SI}). For different positions $x_\textrm{LE}$, the simulation results are used to calculate $S_{21}(\omega,x_\textrm{LE})$, illustrated in Fig.~\ref{fig:Model_1}(a) for $R=0.005\,\mathrm{\Omega}$ according to $Q_\textrm{i}\approx8200$. Regarding the corresponding resonance frequencies $\omega_\textrm{r}(x_\textrm{LE})$ and external quality factors $Q_\textrm{e}(x_\textrm{LE})$ in Fig.~\ref{fig:Model_1}(b), one finds that the external quality factor $Q_\textrm{e}$ diverges at $x_1\approx85\,\mathrm{\mu m}$, where the resonance frequency also has a maximum, whereas at $x_2\approx-65\,\mathrm{\mu m}$ and $x_3\approx230\,\mathrm{\mu m}$ both $\omega_\textrm{r}(x_\textrm{LE})$ and $Q_\textrm{e}(x_\textrm{LE})$ have minima.
Moreover, on the one hand the maximum at $x_\textrm{LE}=x_1$ appears to be more peaked for the external quality factor than for the resonance frequency, on the other hand, the minima ($x_\textrm{LE}=x_{2,3}$) are more spiky for the resonance frequency (note, that this behavior is also found for a linear plot of $Q_\textrm{e}(x_\textrm{LE})$). This behavior enables the discrimination of regimes, in which either shifts of the resonance frequency $\omega_\textrm{r}$ compared to the resonator linewidth $\omega_\textrm{r}/Q_\textrm{L}$ or the relative change of the external quality factor $Q_\textrm{e}$ is dominant. Here, $Q_\textrm{L}$ is the loaded quality factor calculated via $1/Q_\textrm{L}=1/Q_\textrm{e}+1/Q_\textrm{i}$, where $Q_\textrm{i}$ is the internal quality factor. In order to analyze these regimes in more detail, we calculate the relative change of the external quality factor $dQ_\textrm{e}/dx_\textrm{LE}\cdot1/Q_\textrm{e}$, and the spatial derivative $d\omega_\textrm{r}/dx_\textrm{LE}$ of the resonance frequency, normalized to the resonator linewidth $\omega_\textrm{r}/Q_\textrm{L}$ (cf. Fig.~\ref{fig:Model_1}(c)). For the calculations, the internal quality factor was chosen to be $Q_\textrm{i,4\,K}=10^4$ and $Q_\textrm{i,mK}=10^5$ \cite{geerlings2012} to account for internal losses at $T=4\,\textrm{K}$ and $\textrm{mK}$ temperatures, respectively. For an intuitive understanding of the regimes it is useful to look at the absolute values of the resonator properties (Fig.~\ref{fig:Model_1}(b)) and their normalized spatial derivatives (Fig.~\ref{fig:Model_1}(c)) simultaneously. 

In \textbf{regime I} around the maximum at $x_\textrm{LE}=x_1$, the coupling strongly depends on $x_\textrm{LE}$, indicated by the sharp peak of $Q_\textrm{e}$ and the comparatively large values of its relative changes (cf. Fig.~\ref{fig:Model_1}(b), (c)). However, the absolute change of the resonance frequency is rather small. Thus, in this regime, ranging from $60\,\mathrm{\mu m}<x_\textrm{LE}<100\,\mathrm{\mu m}$, the position can be varied such that the device is either over-coupled ($Q_\textrm{e}<Q_\textrm{i}$), or under-coupled ($Q_\textrm{e}>Q_\textrm{i}$), without causing significant changes of the resonance frequency. In particular, at the maximum of the external quality factor ($x_1=85\,\mathrm{\mu m}$), corresponding to the minimum of the mutual inductance $M$, one finds a position where almost no coupling can be observed, which we call "zero coupling position" in the following (cf. Fig.~\ref{fig:Model_1}(a)). 

One further regime, namely \textbf{regime II}, can be found close to the minima $x_{2,3}$ ($-85\,\mathrm{\mu m}<x_\textrm{LE}<-45\,\mathrm{\mu m}$ and $210\,\mathrm{\mu m}<x_\textrm{LE}<250\,\mathrm{\mu m}$). Here, according to the vanishing spatial derivatives, neither of the resonator properties shows significant changes. Thus, for these positions, the resonator properties are quite stable against variations of the position $x_\textrm{LE}$. 

Finally, in \textbf{regime III}, which can be found within $-200\,\mathrm{\mu m}<x_\textrm{LE}<-100\,\mathrm{\mu m}$ and $260\,\mathrm{\mu m}<x_\textrm{LE}<400\,\mathrm{\mu m}$, the relative changes of $Q_\textrm{e}$ are still rather small, however, the resonance frequency is shifted. In this regime, varying the position can be efficiently used to tune the resonance frequency of the device, without leaving the over-coupled limit ($Q_\textrm{e}<Q_\textrm{i}$). Therefore, the model predicts a tunability of $25\,\textrm{MHz}$. Opposed to using SQUID arrays or ferroelectrics, the tuning due to the variation of the position does not result in significant changes of the internal quality factor \cite{palacios2008tunable,sandberg2008tuning,ferdinand2017tunable}. 

In summary, with some restrictions the device enables operation in regimes where both the external quality factor and the resonance frequency can be shifted independently of each other. In addition, there is a regime in the over-coupled limit, where none of the resonator properties depends strongly on variations of the position $x$.

\section{Simulation vs. experiment}
\label{sec:simu}

As a demonstration of suitability, we make a comparison between the model, using the simulation results, and measurement, for varying position $x_\textrm{LE}$. For experimental details see \cite{ferdinand2017SI}.
\begin{figure}[htbp]
		\includegraphics[width=0.47\textwidth]{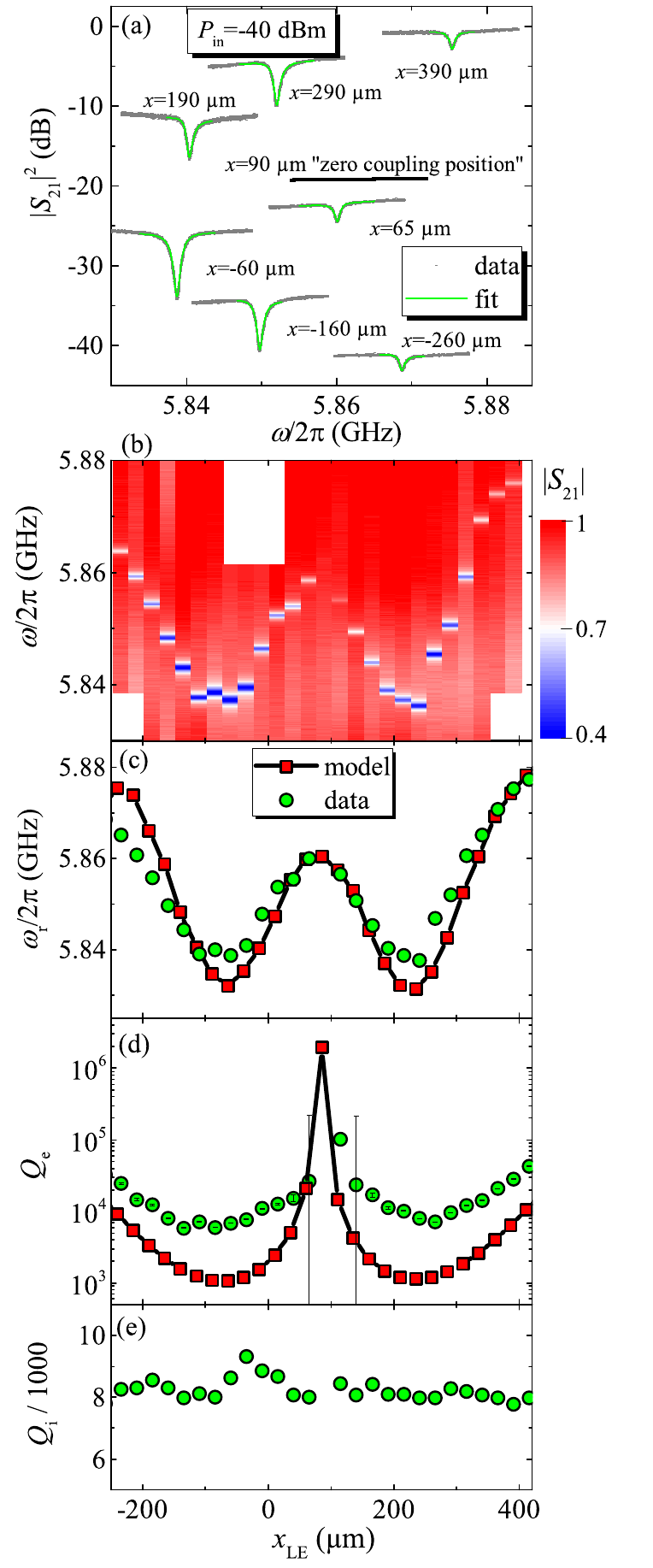}
	\caption{Comparison of model and measurement. (a) transmission measurements for different positions $x_\textrm{LE}$ with an input power of $P_\textrm{in}=-40\,\textrm{dBm}$ (grey) at a temperature of $T=4.2\,\textrm{K}$ with best fits using Eq.~\ref{eq:S21} (green lines) - for clarity, data are shifted along the vertical axis arbitrarily. Black line corresponds to data at the zero coupling position. (b) measured $|S_{21}(\omega,x_\textrm{LE})|$ at $T=4.2\,\textrm{K}$, (c)-(e) $\omega_\textrm{r}$, $Q_\textrm{e}$ and $Q_\textrm{i}$ as a function of $x_\textrm{LE}$ for model (red squares) and measured data (green circles).}
	\label{fig:Res}
\end{figure}
At a temperature of $T=4.2\,\textrm{K}$, where tunneling two-level-systems are saturated, we measure the transmission $S_{21}(\omega,x_\textrm{LE})$ (Fig.~\ref{fig:Res}(a), (b)). Regarding Figs.~\ref{fig:Model_1}(a) and \ref{fig:Res}(b) the calculated and measured forward scattering parameter $S_{21}(\omega,x_\textrm{LE})$ can be compared, demonstrating a qualitatively good agreement. As predicted by the model, by changing the position $x_\textrm{LE}$, the measured resonance frequency of the device is shifted. Furthermore, we observe that the depth of the resonance dip is modified, indicating position dependent coupling strengths. At $x_\textrm{LE}=85\,\mathrm{\mu m}$ there is no resonance dip in both the model and the measurement. More precisely, in accordance with the model, Fig.~\ref{fig:Res}(b) demonstrates that the depth of the resonance dip decreases when $x_\textrm{LE}$ is increased from $-65\,\mathrm{\mu m}$ to $\approx85\,\mathrm{\mu m}$, and gets deeper again when $x_\textrm{LE}$ is increased even more.

We fit the measured data using Eq.~(S10) \cite{ferdinand2017SI,deng2013},
\begin{equation} \label{eq:S21}
S_{21}(\omega)=Ae^{i\phi}\frac{1+2iQ_\textrm{i}\frac{\Delta \omega}{\omega_\textrm{r}}}{1+\frac{Q_\textrm{i}}{Q_\textrm{e}}+i\frac{Q_\textrm{i}}{Q_\textrm{a}}+2iQ_\textrm{i}\frac{\Delta \omega}{\omega_\textrm{r}}}\,\, .
\end{equation}
The dimensionless parameter $Q_\textrm{a}$ accounts for an asymmetric lineshape of the transmission function around the resonance frequency $\omega_\textrm{r}$. For a given resonance curve the errors in $Q_\textrm{e}$, $Q_\textrm{i}$ and $Q_\textrm{a}$ typically are below $1.5\%$, for the resonance frequency they are even below $2.5\cdot10^{-5}\%$. These errors are far below the symbol sizes in Figures~\ref{fig:Res}(c), (d) and (e). Larger errors arise from an uncertainty of $\pm5\,\mathrm{\mu m}$ in the determination of $x_\textrm{LE}$, leading to an error in $\omega_\textrm{r}$ below $0.05\%$, which is comparable to the symbol size. For $Q_\textrm{e}$ the error strongly depends on the position and is indicated in Fig.~\ref{fig:Res}(d) by error bars. Further, we observe run-to-run variations of  $Q_\textrm{e}$ and $Q_\textrm{i}$ of about $5\%$ which is the dominating error source for Qe. We compare the extracted fitting parameters $\omega_\textrm{r}$, $Q_\textrm{e}$ (green circles in Fig.~\ref{fig:Res}(c),(d)) with the calculations of the model using the simulations (red squares). For the sake of comparability, the calculated resonance frequencies were shifted upwards by a factor of $1.0023$. This small difference could originate from a slightly higher substrate thickness of the LE-chip, resulting in smaller capacitances $C_\textrm{c}$, $C_\textrm{s}$ and higher resonance frequencies. Then, model and the measurement agree both qualitatively and quantitatively very well for the resonance frequency $\omega_\textrm{r}$. Despite the qualitative agreement, the external quality factor $Q_\textrm{e}$ is underestimated by the model since it does not include the entire experimental setup, e.g. the in- and output coupling to the SMA connectors (note, that this is the reason for the different color scales in Figs.~\ref{fig:Model_1}(a) and \ref{fig:Res}(b)). This can lead to standing waves on the transmission line \cite{deng2013}, and thereby can decrease the coupling between the resonator and the TL for the resonator being close to a magnetic field node of such a standing wave. One further issue, falsifying the extracted value of $Q_\textrm{e}$, is the finite width of the ground planes of the TL, disabling proper grounding conditions, thereby supporting parasitic transmission through the sample holder \cite{wenner2014catching,bothner2017improving}, which in that case acts as a hollow waveguide. Thus, some parasitic microwave power is transmitted from one port to the other, masking the ratio of off-resonance and on-resonance transmission, which, due to the fit with Eq.~(\ref{eq:S21}), defines $Q_\textrm{i}/Q_\textrm{e}$. However, as the model predicts for $x_\textrm{LE}\approx 85\,\mathrm{\mu m}$, also in the measurements there is a certain position above the signal line of the TL, at which no coupling between resonator and TL can be observed. Such a position corresponds to a huge external quality factor $Q_\textrm{e}$ (peak in Fig.~\ref{fig:Res}(d)). Experimentally, we find a zero coupling position at $x_\textrm{LE}=90\,\mathrm{\mu m}$ (see black data in Fig.~\ref{fig:Res}(b)), which is close to the position predicted by the model. The measurement for this position does not show a resonance dip. Hence, in Fig.~\ref{fig:Res}(c)-(e), at $x_\textrm{LE}=90\,\mathrm{\mu m}$ one data point of the extracted fit parameters of the measurement is missing. As expected, the internal quality factor $Q_\textrm{i}$ does not show a systematic dependency on the position $x_\textrm{LE}$, and is $Q_\textrm{i}\approx8200$ (Fig.~\ref{fig:Res}(e)).

In summary, we have investigated a superconducting microwave device consisting of a lumped element resonator, which is placed on top of a transmission line. We have employed a full model of the circuit and have demonstrated, that the device can be operated in different regimes, which enable control over both the resonance frequency and the coupling between the two subsystems. Furthermore, the device offers the possibility of tuning both the resonance frequency and the coupling to the TL without significantly influencing the internal quality factor of the resonator. \\ 

\section{Acknowledgements}
This work was supported by the DFG via SFB/TRR 21 and the European Research Council via SOCATHES. BF gratefully acknowledges support from the Carl-Zeiss-Stiftung. 

\bibliography{Ferdinand_LE-Model}

\begin{thebibliography}{40}%
\makeatletter
\providecommand \@ifxundefined [1]{%
 \@ifx{#1\undefined}
}%
\providecommand \@ifnum [1]{%
 \ifnum #1\expandafter \@firstoftwo
 \else \expandafter \@secondoftwo
 \fi
}%
\providecommand \@ifx [1]{%
 \ifx #1\expandafter \@firstoftwo
 \else \expandafter \@secondoftwo
 \fi
}%
\providecommand \natexlab [1]{#1}%
\providecommand \enquote  [1]{``#1''}%
\providecommand \bibnamefont  [1]{#1}%
\providecommand \bibfnamefont [1]{#1}%
\providecommand \citenamefont [1]{#1}%
\providecommand \href@noop [0]{\@secondoftwo}%
\providecommand \href [0]{\begingroup \@sanitize@url \@href}%
\providecommand \@href[1]{\@@startlink{#1}\@@href}%
\providecommand \@@href[1]{\endgroup#1\@@endlink}%
\providecommand \@sanitize@url [0]{\catcode `\\12\catcode `\$12\catcode
  `\&12\catcode `\#12\catcode `\^12\catcode `\_12\catcode `\%12\relax}%
\providecommand \@@startlink[1]{}%
\providecommand \@@endlink[0]{}%
\providecommand \url  [0]{\begingroup\@sanitize@url \@url }%
\providecommand \@url [1]{\endgroup\@href {#1}{\urlprefix }}%
\providecommand \urlprefix  [0]{URL }%
\providecommand \Eprint [0]{\href }%
\providecommand \doibase [0]{http://dx.doi.org/}%
\providecommand \selectlanguage [0]{\@gobble}%
\providecommand \bibinfo  [0]{\@secondoftwo}%
\providecommand \bibfield  [0]{\@secondoftwo}%
\providecommand \translation [1]{[#1]}%
\providecommand \BibitemOpen [0]{}%
\providecommand \bibitemStop [0]{}%
\providecommand \bibitemNoStop [0]{.\EOS\space}%
\providecommand \EOS [0]{\spacefactor3000\relax}%
\providecommand \BibitemShut  [1]{\csname bibitem#1\endcsname}%
\let\auto@bib@innerbib\@empty
\bibitem [{\citenamefont {Blais}\ \emph {et~al.}(2004)\citenamefont {Blais},
  \citenamefont {Huang}, \citenamefont {Wallraff}, \citenamefont {Girvin},\
  and\ \citenamefont {Schoelkopf}}]{blais2004cavity}%
  \BibitemOpen
  \bibfield  {author} {\bibinfo {author} {\bibfnamefont {A.}~\bibnamefont
  {Blais}}, \bibinfo {author} {\bibfnamefont {R.-S.}\ \bibnamefont {Huang}},
  \bibinfo {author} {\bibfnamefont {A.}~\bibnamefont {Wallraff}}, \bibinfo
  {author} {\bibfnamefont {S.~M.}\ \bibnamefont {Girvin}}, \ and\ \bibinfo
  {author} {\bibfnamefont {R.~J.}\ \bibnamefont {Schoelkopf}},\ }\bibfield
  {title} {\enquote {\bibinfo {title} {Cavity quantum electrodynamics for
  superconducting electrical circuits: An architecture for quantum
  computation},}\ }\href {\doibase 10.1103/PhysRevA.69.062320} {\bibfield
  {journal} {\bibinfo  {journal} {Phys. Rev. A}\ }\textbf {\bibinfo {volume}
  {69}},\ \bibinfo {pages} {062320} (\bibinfo {year} {2004})}\BibitemShut
  {NoStop}%
\bibitem [{\citenamefont {Wallraff}\ \emph {et~al.}(2004)\citenamefont
  {Wallraff}, \citenamefont {Schuster}, \citenamefont {Blais}, \citenamefont
  {Frunzio}, \citenamefont {Huang}, \citenamefont {Majer}, \citenamefont
  {Kumar}, \citenamefont {Girvin},\ and\ \citenamefont
  {Schoelkopf}}]{wallraff2004strong}%
  \BibitemOpen
  \bibfield  {author} {\bibinfo {author} {\bibfnamefont {A.}~\bibnamefont
  {Wallraff}}, \bibinfo {author} {\bibfnamefont {D.~I.}\ \bibnamefont
  {Schuster}}, \bibinfo {author} {\bibfnamefont {A.}~\bibnamefont {Blais}},
  \bibinfo {author} {\bibfnamefont {L.}~\bibnamefont {Frunzio}}, \bibinfo
  {author} {\bibfnamefont {R.-S.}\ \bibnamefont {Huang}}, \bibinfo {author}
  {\bibfnamefont {J.}~\bibnamefont {Majer}}, \bibinfo {author} {\bibfnamefont
  {S.}~\bibnamefont {Kumar}}, \bibinfo {author} {\bibfnamefont {S.~M.}\
  \bibnamefont {Girvin}}, \ and\ \bibinfo {author} {\bibfnamefont {R.~J.}\
  \bibnamefont {Schoelkopf}},\ }\bibfield  {title} {\enquote {\bibinfo {title}
  {Strong coupling of a single photon to a superconducting qubit using circuit
  quantum electrodynamics},}\ }\href@noop {} {\bibfield  {journal} {\bibinfo
  {journal} {Nature}\ }\textbf {\bibinfo {volume} {431}},\ \bibinfo {pages}
  {162--167} (\bibinfo {year} {2004})}\BibitemShut {NoStop}%
\bibitem [{\citenamefont {Wallraff}\ \emph {et~al.}(2005)\citenamefont
  {Wallraff}, \citenamefont {Schuster}, \citenamefont {Blais}, \citenamefont
  {Frunzio}, \citenamefont {Majer}, \citenamefont {Devoret}, \citenamefont
  {Girvin},\ and\ \citenamefont {Schoelkopf}}]{wallraff2005}%
  \BibitemOpen
  \bibfield  {author} {\bibinfo {author} {\bibfnamefont {A.}~\bibnamefont
  {Wallraff}}, \bibinfo {author} {\bibfnamefont {D.~I.}\ \bibnamefont
  {Schuster}}, \bibinfo {author} {\bibfnamefont {A.}~\bibnamefont {Blais}},
  \bibinfo {author} {\bibfnamefont {L.}~\bibnamefont {Frunzio}}, \bibinfo
  {author} {\bibfnamefont {J.}~\bibnamefont {Majer}}, \bibinfo {author}
  {\bibfnamefont {M.~H.}\ \bibnamefont {Devoret}}, \bibinfo {author}
  {\bibfnamefont {S.~M.}\ \bibnamefont {Girvin}}, \ and\ \bibinfo {author}
  {\bibfnamefont {R.~J.}\ \bibnamefont {Schoelkopf}},\ }\bibfield  {title}
  {\enquote {\bibinfo {title} {Approaching unit visibility for control of a
  superconducting qubit with dispersive readout},}\ }\href@noop {} {\bibfield
  {journal} {\bibinfo  {journal} {Phys. Rev. Lett.}\ }\textbf {\bibinfo
  {volume} {95}},\ \bibinfo {pages} {060501} (\bibinfo {year}
  {2005})}\BibitemShut {NoStop}%
\bibitem [{\citenamefont {Fink}\ \emph {et~al.}(2008)\citenamefont {Fink},
  \citenamefont {G{\"o}ppl}, \citenamefont {Baur}, \citenamefont {Bianchetti},
  \citenamefont {Leek}, \citenamefont {Blais},\ and\ \citenamefont
  {Wallraff}}]{fink2008climbing}%
  \BibitemOpen
  \bibfield  {author} {\bibinfo {author} {\bibfnamefont {J.~M.}\ \bibnamefont
  {Fink}}, \bibinfo {author} {\bibfnamefont {M.}~\bibnamefont {G{\"o}ppl}},
  \bibinfo {author} {\bibfnamefont {M.}~\bibnamefont {Baur}}, \bibinfo {author}
  {\bibfnamefont {R.}~\bibnamefont {Bianchetti}}, \bibinfo {author}
  {\bibfnamefont {P.~J.}\ \bibnamefont {Leek}}, \bibinfo {author}
  {\bibfnamefont {A.}~\bibnamefont {Blais}}, \ and\ \bibinfo {author}
  {\bibfnamefont {A.}~\bibnamefont {Wallraff}},\ }\bibfield  {title} {\enquote
  {\bibinfo {title} {Climbing the {J}aynes--{C}ummings ladder and observing its
  nonlinearity in a cavity {QED} system},}\ }\href@noop {} {\bibfield
  {journal} {\bibinfo  {journal} {Nature}\ }\textbf {\bibinfo {volume} {454}},\
  \bibinfo {pages} {315--318} (\bibinfo {year} {2008})}\BibitemShut {NoStop}%
\bibitem [{\citenamefont {Hofheinz}\ \emph {et~al.}(2009)\citenamefont
  {Hofheinz}, \citenamefont {Wang}, \citenamefont {Ansmann}, \citenamefont
  {Bialczak}, \citenamefont {Lucero}, \citenamefont {Neeley}, \citenamefont
  {O'Connell}, \citenamefont {Sank}, \citenamefont {Wenner}, \citenamefont
  {Martinis},\ and\ \citenamefont {Cleland}}]{hofheinz2009synthesizing}%
  \BibitemOpen
  \bibfield  {author} {\bibinfo {author} {\bibfnamefont {M.}~\bibnamefont
  {Hofheinz}}, \bibinfo {author} {\bibfnamefont {H.}~\bibnamefont {Wang}},
  \bibinfo {author} {\bibfnamefont {M.}~\bibnamefont {Ansmann}}, \bibinfo
  {author} {\bibfnamefont {R.~C.}\ \bibnamefont {Bialczak}}, \bibinfo {author}
  {\bibfnamefont {E.}~\bibnamefont {Lucero}}, \bibinfo {author} {\bibfnamefont
  {M.}~\bibnamefont {Neeley}}, \bibinfo {author} {\bibfnamefont {A.~D.}\
  \bibnamefont {O'Connell}}, \bibinfo {author} {\bibfnamefont {D.}~\bibnamefont
  {Sank}}, \bibinfo {author} {\bibfnamefont {J.}~\bibnamefont {Wenner}},
  \bibinfo {author} {\bibfnamefont {J.~M.}\ \bibnamefont {Martinis}}, \ and\
  \bibinfo {author} {\bibfnamefont {A.}~\bibnamefont {Cleland}},\ }\bibfield
  {title} {\enquote {\bibinfo {title} {Synthesizing arbitrary quantum states in
  a superconducting resonator},}\ }\href@noop {} {\bibfield  {journal}
  {\bibinfo  {journal} {Nature}\ }\textbf {\bibinfo {volume} {459}},\ \bibinfo
  {pages} {546--549} (\bibinfo {year} {2009})}\BibitemShut {NoStop}%
\bibitem [{\citenamefont {Niemczyk}\ \emph {et~al.}(2010)\citenamefont
  {Niemczyk}, \citenamefont {Deppe}, \citenamefont {Huebl}, \citenamefont
  {Menzel}, \citenamefont {Hocke}, \citenamefont {Schwarz}, \citenamefont
  {Garcia-Ripoll}, \citenamefont {Zueco}, \citenamefont {H{\"u}mmer},
  \citenamefont {Solano}, \citenamefont {Marx},\ and\ \citenamefont
  {Gross}}]{niemczyk2010circuit}%
  \BibitemOpen
  \bibfield  {author} {\bibinfo {author} {\bibfnamefont {T.}~\bibnamefont
  {Niemczyk}}, \bibinfo {author} {\bibfnamefont {F.}~\bibnamefont {Deppe}},
  \bibinfo {author} {\bibfnamefont {H.}~\bibnamefont {Huebl}}, \bibinfo
  {author} {\bibfnamefont {E.~P.}\ \bibnamefont {Menzel}}, \bibinfo {author}
  {\bibfnamefont {F.}~\bibnamefont {Hocke}}, \bibinfo {author} {\bibfnamefont
  {M.~J.}\ \bibnamefont {Schwarz}}, \bibinfo {author} {\bibfnamefont {J.~J.}\
  \bibnamefont {Garcia-Ripoll}}, \bibinfo {author} {\bibfnamefont
  {D.}~\bibnamefont {Zueco}}, \bibinfo {author} {\bibfnamefont
  {T.}~\bibnamefont {H{\"u}mmer}}, \bibinfo {author} {\bibfnamefont
  {E.}~\bibnamefont {Solano}}, \bibinfo {author} {\bibfnamefont
  {A.}~\bibnamefont {Marx}}, \ and\ \bibinfo {author} {\bibfnamefont
  {R.}~\bibnamefont {Gross}},\ }\bibfield  {title} {\enquote {\bibinfo {title}
  {Circuit quantum electrodynamics in the ultrastrong-coupling regime},}\
  }\href@noop {} {\bibfield  {journal} {\bibinfo  {journal} {Nature Physics}\
  }\textbf {\bibinfo {volume} {6}},\ \bibinfo {pages} {772--776} (\bibinfo
  {year} {2010})}\BibitemShut {NoStop}%
\bibitem [{\citenamefont {Wilson}\ \emph {et~al.}(2011)\citenamefont {Wilson},
  \citenamefont {Johansson}, \citenamefont {Pourkabirian}, \citenamefont
  {Simoen}, \citenamefont {Johansson}, \citenamefont {Duty}, \citenamefont
  {Nori},\ and\ \citenamefont {Delsing}}]{wilson2011observation}%
  \BibitemOpen
  \bibfield  {author} {\bibinfo {author} {\bibfnamefont {C.~M.}\ \bibnamefont
  {Wilson}}, \bibinfo {author} {\bibfnamefont {G.}~\bibnamefont {Johansson}},
  \bibinfo {author} {\bibfnamefont {A.}~\bibnamefont {Pourkabirian}}, \bibinfo
  {author} {\bibfnamefont {M.}~\bibnamefont {Simoen}}, \bibinfo {author}
  {\bibfnamefont {J.~R.}\ \bibnamefont {Johansson}}, \bibinfo {author}
  {\bibfnamefont {T.}~\bibnamefont {Duty}}, \bibinfo {author} {\bibfnamefont
  {F.}~\bibnamefont {Nori}}, \ and\ \bibinfo {author} {\bibfnamefont
  {P.}~\bibnamefont {Delsing}},\ }\bibfield  {title} {\enquote {\bibinfo
  {title} {Observation of the dynamical {C}asimir effect in a superconducting
  circuit},}\ }\href@noop {} {\bibfield  {journal} {\bibinfo  {journal}
  {Nature}\ }\textbf {\bibinfo {volume} {479}},\ \bibinfo {pages} {376--379}
  (\bibinfo {year} {2011})}\BibitemShut {NoStop}%
\bibitem [{\citenamefont {Kelly}\ \emph {et~al.}(2015)\citenamefont {Kelly},
  \citenamefont {Barends}, \citenamefont {Fowler}, \citenamefont {Megrant},
  \citenamefont {Jeffrey}, \citenamefont {White}, \citenamefont {Sank},
  \citenamefont {Mutus}, \citenamefont {Campbell}, \citenamefont {Chen},
  \citenamefont {Chen}, \citenamefont {Chiaro}, \citenamefont {Dunsworth},
  \citenamefont {Hoi}, \citenamefont {Neill}, \citenamefont {O'Malley},
  \citenamefont {Quintana}, \citenamefont {Roushan}, \citenamefont
  {Vainsencher}, \citenamefont {Wenner}, \citenamefont {Cleland},\ and\
  \citenamefont {Martinis}}]{kelly2015state}%
  \BibitemOpen
  \bibfield  {author} {\bibinfo {author} {\bibfnamefont {J.}~\bibnamefont
  {Kelly}}, \bibinfo {author} {\bibfnamefont {R.}~\bibnamefont {Barends}},
  \bibinfo {author} {\bibfnamefont {A.~G.}\ \bibnamefont {Fowler}}, \bibinfo
  {author} {\bibfnamefont {A.}~\bibnamefont {Megrant}}, \bibinfo {author}
  {\bibfnamefont {E.}~\bibnamefont {Jeffrey}}, \bibinfo {author} {\bibfnamefont
  {T.~C.}\ \bibnamefont {White}}, \bibinfo {author} {\bibfnamefont
  {D.}~\bibnamefont {Sank}}, \bibinfo {author} {\bibfnamefont {J.~Y.}\
  \bibnamefont {Mutus}}, \bibinfo {author} {\bibfnamefont {B.}~\bibnamefont
  {Campbell}}, \bibinfo {author} {\bibfnamefont {Y.}~\bibnamefont {Chen}},
  \bibinfo {author} {\bibfnamefont {Z.}~\bibnamefont {Chen}}, \bibinfo {author}
  {\bibfnamefont {B.}~\bibnamefont {Chiaro}}, \bibinfo {author} {\bibfnamefont
  {A.}~\bibnamefont {Dunsworth}}, \bibinfo {author} {\bibfnamefont {I.-C.}\
  \bibnamefont {Hoi}}, \bibinfo {author} {\bibfnamefont {C.}~\bibnamefont
  {Neill}}, \bibinfo {author} {\bibfnamefont {P.~J.~J.}\ \bibnamefont
  {O'Malley}}, \bibinfo {author} {\bibfnamefont {C.}~\bibnamefont {Quintana}},
  \bibinfo {author} {\bibfnamefont {P.}~\bibnamefont {Roushan}}, \bibinfo
  {author} {\bibfnamefont {A.}~\bibnamefont {Vainsencher}}, \bibinfo {author}
  {\bibfnamefont {J.}~\bibnamefont {Wenner}}, \bibinfo {author} {\bibfnamefont
  {A.}~\bibnamefont {Cleland}}, \ and\ \bibinfo {author} {\bibfnamefont
  {J.~M.}\ \bibnamefont {Martinis}},\ }\bibfield  {title} {\enquote {\bibinfo
  {title} {State preservation by repetitive error detection in a
  superconducting quantum circuit},}\ }\href@noop {} {\bibfield  {journal}
  {\bibinfo  {journal} {Nature}\ }\textbf {\bibinfo {volume} {519}},\ \bibinfo
  {pages} {66--69} (\bibinfo {year} {2015})}\BibitemShut {NoStop}%
\bibitem [{\citenamefont {Bosman}\ \emph
  {et~al.}(2017{\natexlab{a}})\citenamefont {Bosman}, \citenamefont {Gely},
  \citenamefont {Singh}, \citenamefont {Bruno}, \citenamefont {Bothner},\ and\
  \citenamefont {Steele}}]{bosman2017multi}%
  \BibitemOpen
  \bibfield  {author} {\bibinfo {author} {\bibfnamefont {S.~J.}\ \bibnamefont
  {Bosman}}, \bibinfo {author} {\bibfnamefont {M.~F.}\ \bibnamefont {Gely}},
  \bibinfo {author} {\bibfnamefont {V.}~\bibnamefont {Singh}}, \bibinfo
  {author} {\bibfnamefont {A.}~\bibnamefont {Bruno}}, \bibinfo {author}
  {\bibfnamefont {D.}~\bibnamefont {Bothner}}, \ and\ \bibinfo {author}
  {\bibfnamefont {G.~A.}\ \bibnamefont {Steele}},\ }\bibfield  {title}
  {\enquote {\bibinfo {title} {Multi-mode ultra-strong coupling in circuit
  quantum electrodynamics},}\ }\href@noop {} {\bibfield  {journal} {\bibinfo
  {journal} {npj Quantum Information}\ }\textbf {\bibinfo {volume} {3}},\
  \bibinfo {pages} {46} (\bibinfo {year} {2017}{\natexlab{a}})}\BibitemShut
  {NoStop}%
\bibitem [{\citenamefont {Andr{\'e}}\ \emph {et~al.}(2006)\citenamefont
  {Andr{\'e}}, \citenamefont {DeMille}, \citenamefont {Doyle}, \citenamefont
  {Lukin}, \citenamefont {Maxwell}, \citenamefont {Rabl}, \citenamefont
  {Schoelkopf},\ and\ \citenamefont {Zoller}}]{andre2006coherent}%
  \BibitemOpen
  \bibfield  {author} {\bibinfo {author} {\bibfnamefont {A.}~\bibnamefont
  {Andr{\'e}}}, \bibinfo {author} {\bibfnamefont {D.}~\bibnamefont {DeMille}},
  \bibinfo {author} {\bibfnamefont {J.~M.}\ \bibnamefont {Doyle}}, \bibinfo
  {author} {\bibfnamefont {M.~D.}\ \bibnamefont {Lukin}}, \bibinfo {author}
  {\bibfnamefont {S.t~Ex.}\ \bibnamefont {Maxwell}}, \bibinfo {author}
  {\bibfnamefont {P.}~\bibnamefont {Rabl}}, \bibinfo {author} {\bibfnamefont
  {R.~J.}\ \bibnamefont {Schoelkopf}}, \ and\ \bibinfo {author} {\bibfnamefont
  {P.}~\bibnamefont {Zoller}},\ }\bibfield  {title} {\enquote {\bibinfo {title}
  {A coherent all-electrical interface between polar molecules and mesoscopic
  superconducting resonators},}\ }\href@noop {} {\bibfield  {journal} {\bibinfo
   {journal} {Nature Physics}\ }\textbf {\bibinfo {volume} {2}},\ \bibinfo
  {pages} {636--642} (\bibinfo {year} {2006})}\BibitemShut {NoStop}%
\bibitem [{\citenamefont {Rabl}\ \emph {et~al.}(2006)\citenamefont {Rabl},
  \citenamefont {DeMille}, \citenamefont {Doyle}, \citenamefont {Lukin},
  \citenamefont {Schoelkopf},\ and\ \citenamefont {Zoller}}]{rabl2006hybrid}%
  \BibitemOpen
  \bibfield  {author} {\bibinfo {author} {\bibfnamefont {P.}~\bibnamefont
  {Rabl}}, \bibinfo {author} {\bibfnamefont {D.}~\bibnamefont {DeMille}},
  \bibinfo {author} {\bibfnamefont {J.~M.}\ \bibnamefont {Doyle}}, \bibinfo
  {author} {\bibfnamefont {M.~D.}\ \bibnamefont {Lukin}}, \bibinfo {author}
  {\bibfnamefont {R.~J.}\ \bibnamefont {Schoelkopf}}, \ and\ \bibinfo {author}
  {\bibfnamefont {P.}~\bibnamefont {Zoller}},\ }\bibfield  {title} {\enquote
  {\bibinfo {title} {Hybrid quantum processors: molecular ensembles as quantum
  memory for solid state circuits},}\ }\href@noop {} {\bibfield  {journal}
  {\bibinfo  {journal} {Phys. Rev. Lett.}\ }\textbf {\bibinfo {volume} {97}},\
  \bibinfo {pages} {033003} (\bibinfo {year} {2006})}\BibitemShut {NoStop}%
\bibitem [{\citenamefont {Hafezi}\ \emph {et~al.}(2012)\citenamefont {Hafezi},
  \citenamefont {Kim}, \citenamefont {Rolston}, \citenamefont {Orozco},
  \citenamefont {Lev},\ and\ \citenamefont {Taylor}}]{hafezi2012atomic}%
  \BibitemOpen
  \bibfield  {author} {\bibinfo {author} {\bibfnamefont {M.}~\bibnamefont
  {Hafezi}}, \bibinfo {author} {\bibfnamefont {Z.}~\bibnamefont {Kim}},
  \bibinfo {author} {\bibfnamefont {S.~L.}\ \bibnamefont {Rolston}}, \bibinfo
  {author} {\bibfnamefont {L.~A.}\ \bibnamefont {Orozco}}, \bibinfo {author}
  {\bibfnamefont {B.~L.}\ \bibnamefont {Lev}}, \ and\ \bibinfo {author}
  {\bibfnamefont {J.~M.}\ \bibnamefont {Taylor}},\ }\bibfield  {title}
  {\enquote {\bibinfo {title} {Atomic interface between microwave and optical
  photons},}\ }\href@noop {} {\bibfield  {journal} {\bibinfo  {journal} {Phys.
  Rev. A}\ }\textbf {\bibinfo {volume} {85}},\ \bibinfo {pages} {020302}
  (\bibinfo {year} {2012})}\BibitemShut {NoStop}%
\bibitem [{\citenamefont {Petrosyan}\ and\ \citenamefont
  {Fleischhauer}(2008)}]{petrosyan2008quantum}%
  \BibitemOpen
  \bibfield  {author} {\bibinfo {author} {\bibfnamefont {D.}~\bibnamefont
  {Petrosyan}}\ and\ \bibinfo {author} {\bibfnamefont {M.}~\bibnamefont
  {Fleischhauer}},\ }\bibfield  {title} {\enquote {\bibinfo {title} {Quantum
  information processing with single photons and atomic ensembles in microwave
  coplanar waveguide resonators},}\ }\href {\doibase
  10.1103/PhysRevLett.100.170501} {\bibfield  {journal} {\bibinfo  {journal}
  {Phys. Rev. Lett.}\ }\textbf {\bibinfo {volume} {100}},\ \bibinfo {pages}
  {170501} (\bibinfo {year} {2008})}\BibitemShut {NoStop}%
\bibitem [{\citenamefont {Verd{\'u}}\ \emph {et~al.}(2009)\citenamefont
  {Verd{\'u}}, \citenamefont {Zoubi}, \citenamefont {Koller}, \citenamefont
  {Majer}, \citenamefont {Ritsch},\ and\ \citenamefont
  {Schmiedmayer}}]{verdu2009strong}%
  \BibitemOpen
  \bibfield  {author} {\bibinfo {author} {\bibfnamefont {J.}~\bibnamefont
  {Verd{\'u}}}, \bibinfo {author} {\bibfnamefont {H.}~\bibnamefont {Zoubi}},
  \bibinfo {author} {\bibfnamefont {C.}~\bibnamefont {Koller}}, \bibinfo
  {author} {\bibfnamefont {J.}~\bibnamefont {Majer}}, \bibinfo {author}
  {\bibfnamefont {H.}~\bibnamefont {Ritsch}}, \ and\ \bibinfo {author}
  {\bibfnamefont {J.}~\bibnamefont {Schmiedmayer}},\ }\bibfield  {title}
  {\enquote {\bibinfo {title} {Strong magnetic coupling of an ultracold gas to
  a superconducting waveguide cavity},}\ }\href@noop {} {\bibfield  {journal}
  {\bibinfo  {journal} {Phys. Rev. Lett.}\ }\textbf {\bibinfo {volume} {103}},\
  \bibinfo {pages} {043603} (\bibinfo {year} {2009})}\BibitemShut {NoStop}%
\bibitem [{\citenamefont {Imamo\ifmmode~\breve{g}\else
  \u{g}\fi{}lu}(2009)}]{imamouglu2009cavity}%
  \BibitemOpen
  \bibfield  {author} {\bibinfo {author} {\bibfnamefont {A.}~\bibnamefont
  {Imamo\ifmmode~\breve{g}\else \u{g}\fi{}lu}},\ }\bibfield  {title} {\enquote
  {\bibinfo {title} {Cavity {QED} based on collective magnetic dipole coupling:
  Spin ensembles as hybrid two-level systems},}\ }\href {\doibase
  10.1103/PhysRevLett.102.083602} {\bibfield  {journal} {\bibinfo  {journal}
  {Phys. Rev. Lett.}\ }\textbf {\bibinfo {volume} {102}},\ \bibinfo {pages}
  {083602} (\bibinfo {year} {2009})}\BibitemShut {NoStop}%
\bibitem [{\citenamefont {Henschel}\ \emph {et~al.}(2010)\citenamefont
  {Henschel}, \citenamefont {Majer}, \citenamefont {Schmiedmayer},\ and\
  \citenamefont {Ritsch}}]{henschel2010cavity}%
  \BibitemOpen
  \bibfield  {author} {\bibinfo {author} {\bibfnamefont {K.}~\bibnamefont
  {Henschel}}, \bibinfo {author} {\bibfnamefont {J.}~\bibnamefont {Majer}},
  \bibinfo {author} {\bibfnamefont {J.}~\bibnamefont {Schmiedmayer}}, \ and\
  \bibinfo {author} {\bibfnamefont {H.}~\bibnamefont {Ritsch}},\ }\bibfield
  {title} {\enquote {\bibinfo {title} {Cavity {QED} with an ultracold ensemble
  on a chip: Prospects for strong magnetic coupling at finite temperatures},}\
  }\href {\doibase 10.1103/PhysRevA.82.033810} {\bibfield  {journal} {\bibinfo
  {journal} {Phys. Rev. A}\ }\textbf {\bibinfo {volume} {82}},\ \bibinfo
  {pages} {033810} (\bibinfo {year} {2010})}\BibitemShut {NoStop}%
\bibitem [{\citenamefont {Schuster}\ \emph {et~al.}(2010)\citenamefont
  {Schuster}, \citenamefont {Sears}, \citenamefont {Ginossar}, \citenamefont
  {DiCarlo}, \citenamefont {Frunzio}, \citenamefont {Morton}, \citenamefont
  {Wu}, \citenamefont {Briggs}, \citenamefont {Buckley}, \citenamefont
  {Awschalom},\ and\ \citenamefont {Schoelkopf}}]{schuster2010high}%
  \BibitemOpen
  \bibfield  {author} {\bibinfo {author} {\bibfnamefont {D.~I.}\ \bibnamefont
  {Schuster}}, \bibinfo {author} {\bibfnamefont {A.~P.}\ \bibnamefont {Sears}},
  \bibinfo {author} {\bibfnamefont {E.}~\bibnamefont {Ginossar}}, \bibinfo
  {author} {\bibfnamefont {L.}~\bibnamefont {DiCarlo}}, \bibinfo {author}
  {\bibfnamefont {L.}~\bibnamefont {Frunzio}}, \bibinfo {author} {\bibfnamefont
  {J.~J.~L.}\ \bibnamefont {Morton}}, \bibinfo {author} {\bibfnamefont
  {H.}~\bibnamefont {Wu}}, \bibinfo {author} {\bibfnamefont {G.~A.~D.}\
  \bibnamefont {Briggs}}, \bibinfo {author} {\bibfnamefont {B.~B.}\
  \bibnamefont {Buckley}}, \bibinfo {author} {\bibfnamefont {D.~D.}\
  \bibnamefont {Awschalom}}, \ and\ \bibinfo {author} {\bibfnamefont {R.~J.}\
  \bibnamefont {Schoelkopf}},\ }\bibfield  {title} {\enquote {\bibinfo {title}
  {High-cooperativity coupling of electron-spin ensembles to superconducting
  cavities},}\ }\href {\doibase 10.1103/PhysRevLett.105.140501} {\bibfield
  {journal} {\bibinfo  {journal} {Phys. Rev. Lett.}\ }\textbf {\bibinfo
  {volume} {105}},\ \bibinfo {pages} {140501} (\bibinfo {year}
  {2010})}\BibitemShut {NoStop}%
\bibitem [{\citenamefont {Kubo}\ \emph {et~al.}(2010)\citenamefont {Kubo},
  \citenamefont {Ong}, \citenamefont {Bertet}, \citenamefont {Vion},
  \citenamefont {Jacques}, \citenamefont {Zheng}, \citenamefont {Dr\'eau},
  \citenamefont {Roch}, \citenamefont {Auffeves}, \citenamefont {Jelezko},
  \citenamefont {Wrachtrup}, \citenamefont {Barthe}, \citenamefont {Bergonzo},\
  and\ \citenamefont {Esteve}}]{kubo2010strong}%
  \BibitemOpen
  \bibfield  {author} {\bibinfo {author} {\bibfnamefont {Y.}~\bibnamefont
  {Kubo}}, \bibinfo {author} {\bibfnamefont {F.~R.}\ \bibnamefont {Ong}},
  \bibinfo {author} {\bibfnamefont {P.}~\bibnamefont {Bertet}}, \bibinfo
  {author} {\bibfnamefont {D.}~\bibnamefont {Vion}}, \bibinfo {author}
  {\bibfnamefont {V.}~\bibnamefont {Jacques}}, \bibinfo {author} {\bibfnamefont
  {D.}~\bibnamefont {Zheng}}, \bibinfo {author} {\bibfnamefont
  {A.}~\bibnamefont {Dr\'eau}}, \bibinfo {author} {\bibfnamefont {J.-F.}\
  \bibnamefont {Roch}}, \bibinfo {author} {\bibfnamefont {A.}~\bibnamefont
  {Auffeves}}, \bibinfo {author} {\bibfnamefont {F.}~\bibnamefont {Jelezko}},
  \bibinfo {author} {\bibfnamefont {J.}~\bibnamefont {Wrachtrup}}, \bibinfo
  {author} {\bibfnamefont {M.~F.}\ \bibnamefont {Barthe}}, \bibinfo {author}
  {\bibfnamefont {P.}~\bibnamefont {Bergonzo}}, \ and\ \bibinfo {author}
  {\bibfnamefont {D.}~\bibnamefont {Esteve}},\ }\bibfield  {title} {\enquote
  {\bibinfo {title} {Strong coupling of a spin ensemble to a superconducting
  resonator},}\ }\href {\doibase 10.1103/PhysRevLett.105.140502} {\bibfield
  {journal} {\bibinfo  {journal} {Phys. Rev. Lett.}\ }\textbf {\bibinfo
  {volume} {105}},\ \bibinfo {pages} {140502} (\bibinfo {year}
  {2010})}\BibitemShut {NoStop}%
\bibitem [{\citenamefont {Wu}\ \emph {et~al.}(2010)\citenamefont {Wu},
  \citenamefont {George}, \citenamefont {Wesenberg}, \citenamefont {M{\o}lmer},
  \citenamefont {Schuster}, \citenamefont {Schoelkopf}, \citenamefont {Itoh},
  \citenamefont {Ardavan}, \citenamefont {Morton},\ and\ \citenamefont
  {Briggs}}]{wu2010storage}%
  \BibitemOpen
  \bibfield  {author} {\bibinfo {author} {\bibfnamefont {H.}~\bibnamefont
  {Wu}}, \bibinfo {author} {\bibfnamefont {R.~E.}\ \bibnamefont {George}},
  \bibinfo {author} {\bibfnamefont {J.~H.}\ \bibnamefont {Wesenberg}}, \bibinfo
  {author} {\bibfnamefont {K.}~\bibnamefont {M{\o}lmer}}, \bibinfo {author}
  {\bibfnamefont {D.~I.}\ \bibnamefont {Schuster}}, \bibinfo {author}
  {\bibfnamefont {R.~J.}\ \bibnamefont {Schoelkopf}}, \bibinfo {author}
  {\bibfnamefont {K.~M.}\ \bibnamefont {Itoh}}, \bibinfo {author}
  {\bibfnamefont {A.}~\bibnamefont {Ardavan}}, \bibinfo {author} {\bibfnamefont
  {J.~J.~L.}\ \bibnamefont {Morton}}, \ and\ \bibinfo {author} {\bibfnamefont
  {G.~A.~D.}\ \bibnamefont {Briggs}},\ }\bibfield  {title} {\enquote {\bibinfo
  {title} {Storage of multiple coherent microwave excitations in an electron
  spin ensemble},}\ }\href@noop {} {\bibfield  {journal} {\bibinfo  {journal}
  {Phys. Rev. Lett.}\ }\textbf {\bibinfo {volume} {105}},\ \bibinfo {pages}
  {140503} (\bibinfo {year} {2010})}\BibitemShut {NoStop}%
\bibitem [{\citenamefont {Ams\"uss}\ \emph {et~al.}(2011)\citenamefont
  {Ams\"uss}, \citenamefont {Koller}, \citenamefont {N\"obauer}, \citenamefont
  {Putz}, \citenamefont {Rotter}, \citenamefont {Sandner}, \citenamefont
  {Schneider}, \citenamefont {Schramb\"ock}, \citenamefont {Steinhauser},
  \citenamefont {Ritsch}, \citenamefont {Schmiedmayer},\ and\ \citenamefont
  {Majer}}]{amsuss2011cavity}%
  \BibitemOpen
  \bibfield  {author} {\bibinfo {author} {\bibfnamefont {R.}~\bibnamefont
  {Ams\"uss}}, \bibinfo {author} {\bibfnamefont {C.}~\bibnamefont {Koller}},
  \bibinfo {author} {\bibfnamefont {T.}~\bibnamefont {N\"obauer}}, \bibinfo
  {author} {\bibfnamefont {S.}~\bibnamefont {Putz}}, \bibinfo {author}
  {\bibfnamefont {S.}~\bibnamefont {Rotter}}, \bibinfo {author} {\bibfnamefont
  {K.}~\bibnamefont {Sandner}}, \bibinfo {author} {\bibfnamefont
  {S.}~\bibnamefont {Schneider}}, \bibinfo {author} {\bibfnamefont
  {M.}~\bibnamefont {Schramb\"ock}}, \bibinfo {author} {\bibfnamefont
  {G.}~\bibnamefont {Steinhauser}}, \bibinfo {author} {\bibfnamefont
  {H.}~\bibnamefont {Ritsch}}, \bibinfo {author} {\bibfnamefont
  {J.}~\bibnamefont {Schmiedmayer}}, \ and\ \bibinfo {author} {\bibfnamefont
  {J.}~\bibnamefont {Majer}},\ }\bibfield  {title} {\enquote {\bibinfo {title}
  {Cavity {QED} with magnetically coupled collective spin states},}\ }\href
  {\doibase 10.1103/PhysRevLett.107.060502} {\bibfield  {journal} {\bibinfo
  {journal} {Phys. Rev. Lett.}\ }\textbf {\bibinfo {volume} {107}},\ \bibinfo
  {pages} {060502} (\bibinfo {year} {2011})}\BibitemShut {NoStop}%
\bibitem [{\citenamefont {Probst}\ \emph {et~al.}(2013)\citenamefont {Probst},
  \citenamefont {Rotzinger}, \citenamefont {W\"unsch}, \citenamefont {Jung},
  \citenamefont {Jerger}, \citenamefont {Siegel}, \citenamefont {Ustinov},\
  and\ \citenamefont {Bushev}}]{probst2013anisotropic}%
  \BibitemOpen
  \bibfield  {author} {\bibinfo {author} {\bibfnamefont {S.}~\bibnamefont
  {Probst}}, \bibinfo {author} {\bibfnamefont {H.}~\bibnamefont {Rotzinger}},
  \bibinfo {author} {\bibfnamefont {S.}~\bibnamefont {W\"unsch}}, \bibinfo
  {author} {\bibfnamefont {P.}~\bibnamefont {Jung}}, \bibinfo {author}
  {\bibfnamefont {M.}~\bibnamefont {Jerger}}, \bibinfo {author} {\bibfnamefont
  {M.}~\bibnamefont {Siegel}}, \bibinfo {author} {\bibfnamefont {A.~V.}\
  \bibnamefont {Ustinov}}, \ and\ \bibinfo {author} {\bibfnamefont {P.~A.}\
  \bibnamefont {Bushev}},\ }\bibfield  {title} {\enquote {\bibinfo {title}
  {Anisotropic rare-earth spin ensemble strongly coupled to a superconducting
  resonator},}\ }\href {\doibase 10.1103/PhysRevLett.110.157001} {\bibfield
  {journal} {\bibinfo  {journal} {Phys. Rev. Lett.}\ }\textbf {\bibinfo
  {volume} {110}},\ \bibinfo {pages} {157001} (\bibinfo {year}
  {2013})}\BibitemShut {NoStop}%
\bibitem [{\citenamefont {Zollitsch}\ \emph {et~al.}(2015)\citenamefont
  {Zollitsch}, \citenamefont {Mueller}, \citenamefont {Franke}, \citenamefont
  {Goennenwein}, \citenamefont {Brandt}, \citenamefont {Gross},\ and\
  \citenamefont {Huebl}}]{zollitsch2015high}%
  \BibitemOpen
  \bibfield  {author} {\bibinfo {author} {\bibfnamefont {C.~W.}\ \bibnamefont
  {Zollitsch}}, \bibinfo {author} {\bibfnamefont {K.}~\bibnamefont {Mueller}},
  \bibinfo {author} {\bibfnamefont {D.~P.}\ \bibnamefont {Franke}}, \bibinfo
  {author} {\bibfnamefont {S.~T.~B.}\ \bibnamefont {Goennenwein}}, \bibinfo
  {author} {\bibfnamefont {M.~S.}\ \bibnamefont {Brandt}}, \bibinfo {author}
  {\bibfnamefont {R.}~\bibnamefont {Gross}}, \ and\ \bibinfo {author}
  {\bibfnamefont {H.}~\bibnamefont {Huebl}},\ }\bibfield  {title} {\enquote
  {\bibinfo {title} {High cooperativity coupling between a phosphorus donor
  spin ensemble and a superconducting microwave resonator},}\ }\href@noop {}
  {\bibfield  {journal} {\bibinfo  {journal} {Appl. Phys. Lett.}\ }\textbf
  {\bibinfo {volume} {107}},\ \bibinfo {pages} {142105} (\bibinfo {year}
  {2015})}\BibitemShut {NoStop}%
\bibitem [{\citenamefont {Bernon}\ \emph {et~al.}(2013)\citenamefont {Bernon},
  \citenamefont {Hattermann}, \citenamefont {Bothner}, \citenamefont
  {Knufinke}, \citenamefont {Weiss}, \citenamefont {Jessen}, \citenamefont
  {Cano}, \citenamefont {Kemmler}, \citenamefont {Kleiner}, \citenamefont
  {Koelle},\ and\ \citenamefont {Fort\'{a}gh}}]{bernon2013manipulation}%
  \BibitemOpen
  \bibfield  {author} {\bibinfo {author} {\bibfnamefont {S.}~\bibnamefont
  {Bernon}}, \bibinfo {author} {\bibfnamefont {H.}~\bibnamefont {Hattermann}},
  \bibinfo {author} {\bibfnamefont {D.}~\bibnamefont {Bothner}}, \bibinfo
  {author} {\bibfnamefont {M.}~\bibnamefont {Knufinke}}, \bibinfo {author}
  {\bibfnamefont {P.}~\bibnamefont {Weiss}}, \bibinfo {author} {\bibfnamefont
  {F.}~\bibnamefont {Jessen}}, \bibinfo {author} {\bibfnamefont {D.l}\
  \bibnamefont {Cano}}, \bibinfo {author} {\bibfnamefont {M.}~\bibnamefont
  {Kemmler}}, \bibinfo {author} {\bibfnamefont {R.}~\bibnamefont {Kleiner}},
  \bibinfo {author} {\bibfnamefont {D.}~\bibnamefont {Koelle}}, \ and\ \bibinfo
  {author} {\bibfnamefont {J.}~\bibnamefont {Fort\'{a}gh}},\ }\bibfield
  {title} {\enquote {\bibinfo {title} {Manipulation and coherence of ultra-cold
  atoms on a superconducting atom chip},}\ }\href@noop {} {\bibfield  {journal}
  {\bibinfo  {journal} {Nature Commun.}\ }\textbf {\bibinfo {volume} {4}}
  (\bibinfo {year} {2013})}\BibitemShut {NoStop}%
\bibitem [{\citenamefont {Hattermann}\ \emph {et~al.}(2017)\citenamefont
  {Hattermann}, \citenamefont {Bothner}, \citenamefont {Ley}, \citenamefont
  {Ferdinand}, \citenamefont {Wiedmaier}, \citenamefont {S\'{a}rk\'{a}ny},
  \citenamefont {Kleiner}, \citenamefont {Koelle},\ and\ \citenamefont
  {Fort\'{a}gh}}]{Hattermann2017Coupling}%
  \BibitemOpen
  \bibfield  {author} {\bibinfo {author} {\bibfnamefont {H.}~\bibnamefont
  {Hattermann}}, \bibinfo {author} {\bibfnamefont {D.}~\bibnamefont {Bothner}},
  \bibinfo {author} {\bibfnamefont {L.~Y.}\ \bibnamefont {Ley}}, \bibinfo
  {author} {\bibfnamefont {B.}~\bibnamefont {Ferdinand}}, \bibinfo {author}
  {\bibfnamefont {D.}~\bibnamefont {Wiedmaier}}, \bibinfo {author}
  {\bibfnamefont {L.}~\bibnamefont {S\'{a}rk\'{a}ny}}, \bibinfo {author}
  {\bibfnamefont {R.}~\bibnamefont {Kleiner}}, \bibinfo {author} {\bibfnamefont
  {D.}~\bibnamefont {Koelle}}, \ and\ \bibinfo {author} {\bibfnamefont
  {J.}~\bibnamefont {Fort\'{a}gh}},\ }\bibfield  {title} {\enquote {\bibinfo
  {title} {Coupling ultracold atoms to a superconducting coplanar waveguide
  resonator},}\ }\href@noop {} {\bibfield  {journal} {\bibinfo  {journal}
  {Nature Commun.}\ }\textbf {\bibinfo {volume} {8}} (\bibinfo {year}
  {2017})}\BibitemShut {NoStop}%
\bibitem [{\citenamefont {Samkharadze}\ \emph {et~al.}(2016)\citenamefont
  {Samkharadze}, \citenamefont {Bruno}, \citenamefont {Scarlino}, \citenamefont
  {Zheng}, \citenamefont {DiVincenzo}, \citenamefont {DiCarlo},\ and\
  \citenamefont {Vandersypen}}]{samkharadze2016high}%
  \BibitemOpen
  \bibfield  {author} {\bibinfo {author} {\bibfnamefont {N.}~\bibnamefont
  {Samkharadze}}, \bibinfo {author} {\bibfnamefont {A.}~\bibnamefont {Bruno}},
  \bibinfo {author} {\bibfnamefont {P.}~\bibnamefont {Scarlino}}, \bibinfo
  {author} {\bibfnamefont {G.}~\bibnamefont {Zheng}}, \bibinfo {author}
  {\bibfnamefont {D.~P.}\ \bibnamefont {DiVincenzo}}, \bibinfo {author}
  {\bibfnamefont {L.}~\bibnamefont {DiCarlo}}, \ and\ \bibinfo {author}
  {\bibfnamefont {L.~M.~K.}\ \bibnamefont {Vandersypen}},\ }\bibfield  {title}
  {\enquote {\bibinfo {title} {High-kinetic-inductance superconducting nanowire
  resonators for circuit {QED} in a magnetic field},}\ }\href@noop {}
  {\bibfield  {journal} {\bibinfo  {journal} {Phys. Rev. Applied}\ }\textbf
  {\bibinfo {volume} {5}},\ \bibinfo {pages} {044004} (\bibinfo {year}
  {2016})}\BibitemShut {NoStop}%
\bibitem [{\citenamefont {Bienfait}\ \emph {et~al.}(2016)\citenamefont
  {Bienfait}, \citenamefont {Pla}, \citenamefont {Kubo}, \citenamefont {Zhou},
  \citenamefont {Stern}, \citenamefont {Lo}, \citenamefont {Weis},
  \citenamefont {Schenkel}, \citenamefont {Vion}, \citenamefont {Esteve},
  \citenamefont {Morton},\ and\ \citenamefont
  {Bertet}}]{bienfait2016controlling}%
  \BibitemOpen
  \bibfield  {author} {\bibinfo {author} {\bibfnamefont {A.}~\bibnamefont
  {Bienfait}}, \bibinfo {author} {\bibfnamefont {J.~J.}\ \bibnamefont {Pla}},
  \bibinfo {author} {\bibfnamefont {Y.}~\bibnamefont {Kubo}}, \bibinfo {author}
  {\bibfnamefont {X.}~\bibnamefont {Zhou}}, \bibinfo {author} {\bibfnamefont
  {M.}~\bibnamefont {Stern}}, \bibinfo {author} {\bibfnamefont {C.~C.}\
  \bibnamefont {Lo}}, \bibinfo {author} {\bibfnamefont {C.~D.}\ \bibnamefont
  {Weis}}, \bibinfo {author} {\bibfnamefont {T.}~\bibnamefont {Schenkel}},
  \bibinfo {author} {\bibfnamefont {D.}~\bibnamefont {Vion}}, \bibinfo {author}
  {\bibfnamefont {D.}~\bibnamefont {Esteve}}, \bibinfo {author} {\bibfnamefont
  {J.~J.}\ \bibnamefont {Morton}}, \ and\ \bibinfo {author} {\bibfnamefont
  {P.}~\bibnamefont {Bertet}},\ }\bibfield  {title} {\enquote {\bibinfo {title}
  {Controlling spin relaxation with a cavity},}\ }\href@noop {} {\bibfield
  {journal} {\bibinfo  {journal} {Nature}\ } (\bibinfo {year}
  {2016})}\BibitemShut {NoStop}%
\bibitem [{\citenamefont {Sarabi}\ \emph {et~al.}(2017)\citenamefont {Sarabi},
  \citenamefont {Huang},\ and\ \citenamefont {Zimmerman}}]{sarabi2017}%
  \BibitemOpen
  \bibfield  {author} {\bibinfo {author} {\bibfnamefont {B.}~\bibnamefont
  {Sarabi}}, \bibinfo {author} {\bibfnamefont {P.}~\bibnamefont {Huang}}, \
  and\ \bibinfo {author} {\bibfnamefont {N.~M.}\ \bibnamefont {Zimmerman}},\
  }\bibfield  {title} {\enquote {\bibinfo {title} {Prospective two orders of
  magnitude enhancement in direct magnetic coupling of a single-atom spin to a
  circuit resonator},}\ }\href@noop {} {\bibfield  {journal} {\bibinfo
  {journal} {arXiv:1702.02210}\ } (\bibinfo {year} {2017})}\BibitemShut
  {NoStop}%
\bibitem [{\citenamefont {Stockklauser}\ \emph {et~al.}(2017)\citenamefont
  {Stockklauser}, \citenamefont {Scarlino}, \citenamefont {Koski},
  \citenamefont {Gasparinetti}, \citenamefont {Andersen}, \citenamefont
  {Reichl}, \citenamefont {Wegscheider}, \citenamefont {Ihn}, \citenamefont
  {Ensslin},\ and\ \citenamefont {Wallraff}}]{stockklauser2017strong}%
  \BibitemOpen
  \bibfield  {author} {\bibinfo {author} {\bibfnamefont {A.}~\bibnamefont
  {Stockklauser}}, \bibinfo {author} {\bibfnamefont {P.}~\bibnamefont
  {Scarlino}}, \bibinfo {author} {\bibfnamefont {J.~V.}\ \bibnamefont {Koski}},
  \bibinfo {author} {\bibfnamefont {S.}~\bibnamefont {Gasparinetti}}, \bibinfo
  {author} {\bibfnamefont {C.~K.}\ \bibnamefont {Andersen}}, \bibinfo {author}
  {\bibfnamefont {C.}~\bibnamefont {Reichl}}, \bibinfo {author} {\bibfnamefont
  {W.}~\bibnamefont {Wegscheider}}, \bibinfo {author} {\bibfnamefont
  {T.}~\bibnamefont {Ihn}}, \bibinfo {author} {\bibfnamefont {K.}~\bibnamefont
  {Ensslin}}, \ and\ \bibinfo {author} {\bibfnamefont {A.}~\bibnamefont
  {Wallraff}},\ }\bibfield  {title} {\enquote {\bibinfo {title} {Strong
  coupling cavity {QED} with gate-defined double quantum dots enabled by a high
  impedance resonator},}\ }\href@noop {} {\bibfield  {journal} {\bibinfo
  {journal} {Phys. Rev. X}\ }\textbf {\bibinfo {volume} {7}},\ \bibinfo {pages}
  {011030} (\bibinfo {year} {2017})}\BibitemShut {NoStop}%
\bibitem [{\citenamefont {Bosman}\ \emph
  {et~al.}(2017{\natexlab{b}})\citenamefont {Bosman}, \citenamefont {Gely},
  \citenamefont {Singh}, \citenamefont {Bothner}, \citenamefont
  {Castellanos-Gomez},\ and\ \citenamefont {Steele}}]{bosman2017approaching}%
  \BibitemOpen
  \bibfield  {author} {\bibinfo {author} {\bibfnamefont {S.~J.}\ \bibnamefont
  {Bosman}}, \bibinfo {author} {\bibfnamefont {M.~F.}\ \bibnamefont {Gely}},
  \bibinfo {author} {\bibfnamefont {V.}~\bibnamefont {Singh}}, \bibinfo
  {author} {\bibfnamefont {D.}~\bibnamefont {Bothner}}, \bibinfo {author}
  {\bibfnamefont {A.}~\bibnamefont {Castellanos-Gomez}}, \ and\ \bibinfo
  {author} {\bibfnamefont {G.~A.}\ \bibnamefont {Steele}},\ }\bibfield  {title}
  {\enquote {\bibinfo {title} {Approaching ultrastrong coupling in transmon
  circuit {QED} using a high-impedance resonator},}\ }\href {\doibase
  10.1103/PhysRevB.95.224515} {\bibfield  {journal} {\bibinfo  {journal} {Phys.
  Rev. B}\ }\textbf {\bibinfo {volume} {95}},\ \bibinfo {pages} {224515}
  (\bibinfo {year} {2017}{\natexlab{b}})}\BibitemShut {NoStop}%
\bibitem [{\citenamefont {Geerlings}\ \emph {et~al.}(2012)\citenamefont
  {Geerlings}, \citenamefont {Shankar}, \citenamefont {Edwards}, \citenamefont
  {Frunzio}, \citenamefont {Schoelkopf},\ and\ \citenamefont
  {Devoret}}]{geerlings2012}%
  \BibitemOpen
  \bibfield  {author} {\bibinfo {author} {\bibfnamefont {K.}~\bibnamefont
  {Geerlings}}, \bibinfo {author} {\bibfnamefont {S.}~\bibnamefont {Shankar}},
  \bibinfo {author} {\bibfnamefont {E.}~\bibnamefont {Edwards}}, \bibinfo
  {author} {\bibfnamefont {L.}~\bibnamefont {Frunzio}}, \bibinfo {author}
  {\bibfnamefont {R.~J.}\ \bibnamefont {Schoelkopf}}, \ and\ \bibinfo {author}
  {\bibfnamefont {M.~H.}\ \bibnamefont {Devoret}},\ }\bibfield  {title}
  {\enquote {\bibinfo {title} {Improving the quality factor of microwave
  compact resonators by optimizing their geometrical parameters},}\ }\href@noop
  {} {\bibfield  {journal} {\bibinfo  {journal} {Appl. Phys. Lett.}\ }\textbf
  {\bibinfo {volume} {100}},\ \bibinfo {pages} {192601} (\bibinfo {year}
  {2012})}\BibitemShut {NoStop}%
\bibitem [{\citenamefont {Deng}\ \emph {et~al.}(2013)\citenamefont {Deng},
  \citenamefont {Otto},\ and\ \citenamefont {Lupascu}}]{deng2013}%
  \BibitemOpen
  \bibfield  {author} {\bibinfo {author} {\bibfnamefont {C.}~\bibnamefont
  {Deng}}, \bibinfo {author} {\bibfnamefont {M.}~\bibnamefont {Otto}}, \ and\
  \bibinfo {author} {\bibfnamefont {A.}~\bibnamefont {Lupascu}},\ }\bibfield
  {title} {\enquote {\bibinfo {title} {An analysis method for transmission
  measurements of superconducting resonators with applications to
  quantum-regime dielectric-loss measurements},}\ }\href@noop {} {\bibfield
  {journal} {\bibinfo  {journal} {J. Appl. Phys.}\ }\textbf {\bibinfo {volume}
  {114}},\ \bibinfo {pages} {054504} (\bibinfo {year} {2013})}\BibitemShut
  {NoStop}%
\bibitem [{\citenamefont {Palacios-Laloy}\ \emph {et~al.}(2008)\citenamefont
  {Palacios-Laloy}, \citenamefont {Nguyen}, \citenamefont {Mallet},
  \citenamefont {Bertet}, \citenamefont {Vion},\ and\ \citenamefont
  {Esteve}}]{palacios2008tunable}%
  \BibitemOpen
  \bibfield  {author} {\bibinfo {author} {\bibfnamefont {A.}~\bibnamefont
  {Palacios-Laloy}}, \bibinfo {author} {\bibfnamefont {F.}~\bibnamefont
  {Nguyen}}, \bibinfo {author} {\bibfnamefont {F.}~\bibnamefont {Mallet}},
  \bibinfo {author} {\bibfnamefont {P.}~\bibnamefont {Bertet}}, \bibinfo
  {author} {\bibfnamefont {D.}~\bibnamefont {Vion}}, \ and\ \bibinfo {author}
  {\bibfnamefont {D.}~\bibnamefont {Esteve}},\ }\bibfield  {title} {\enquote
  {\bibinfo {title} {Tunable resonators for quantum circuits},}\ }\href@noop {}
  {\bibfield  {journal} {\bibinfo  {journal} {J. Low Temp. Phys.}\ }\textbf
  {\bibinfo {volume} {151}},\ \bibinfo {pages} {1034--1042} (\bibinfo {year}
  {2008})}\BibitemShut {NoStop}%
\bibitem [{\citenamefont {Sandberg}\ \emph {et~al.}(2008)\citenamefont
  {Sandberg}, \citenamefont {Wilson}, \citenamefont {Persson}, \citenamefont
  {Bauch}, \citenamefont {Johansson}, \citenamefont {Shumeiko}, \citenamefont
  {Duty},\ and\ \citenamefont {Delsing}}]{sandberg2008tuning}%
  \BibitemOpen
  \bibfield  {author} {\bibinfo {author} {\bibfnamefont {M.}~\bibnamefont
  {Sandberg}}, \bibinfo {author} {\bibfnamefont {C.~M.}\ \bibnamefont
  {Wilson}}, \bibinfo {author} {\bibfnamefont {F.}~\bibnamefont {Persson}},
  \bibinfo {author} {\bibfnamefont {T.}~\bibnamefont {Bauch}}, \bibinfo
  {author} {\bibfnamefont {G.}~\bibnamefont {Johansson}}, \bibinfo {author}
  {\bibfnamefont {V.}~\bibnamefont {Shumeiko}}, \bibinfo {author}
  {\bibfnamefont {T.}~\bibnamefont {Duty}}, \ and\ \bibinfo {author}
  {\bibfnamefont {P.}~\bibnamefont {Delsing}},\ }\bibfield  {title} {\enquote
  {\bibinfo {title} {Tuning the field in a microwave resonator faster than the
  photon lifetime},}\ }\href@noop {} {\bibfield  {journal} {\bibinfo  {journal}
  {Appl. Phys. Lett.}\ }\textbf {\bibinfo {volume} {92}},\ \bibinfo {pages}
  {203501} (\bibinfo {year} {2008})}\BibitemShut {NoStop}%
\bibitem [{\citenamefont {Vissers}\ \emph {et~al.}(2015)\citenamefont
  {Vissers}, \citenamefont {Hubmayr}, \citenamefont {Sandberg}, \citenamefont
  {Chaudhuri}, \citenamefont {Bockstiegel},\ and\ \citenamefont
  {Gao}}]{vissers2015frequency}%
  \BibitemOpen
  \bibfield  {author} {\bibinfo {author} {\bibfnamefont {M.~R.}\ \bibnamefont
  {Vissers}}, \bibinfo {author} {\bibfnamefont {J.}~\bibnamefont {Hubmayr}},
  \bibinfo {author} {\bibfnamefont {M.}~\bibnamefont {Sandberg}}, \bibinfo
  {author} {\bibfnamefont {S.}~\bibnamefont {Chaudhuri}}, \bibinfo {author}
  {\bibfnamefont {C.}~\bibnamefont {Bockstiegel}}, \ and\ \bibinfo {author}
  {\bibfnamefont {J.}~\bibnamefont {Gao}},\ }\bibfield  {title} {\enquote
  {\bibinfo {title} {Frequency-tunable superconducting resonators via nonlinear
  kinetic inductance},}\ }\href@noop {} {\bibfield  {journal} {\bibinfo
  {journal} {Appl. Phys. Lett.}\ }\textbf {\bibinfo {volume} {107}},\ \bibinfo
  {pages} {062601} (\bibinfo {year} {2015})}\BibitemShut {NoStop}%
\bibitem [{fer()}]{ferdinand2017SI}%
  \BibitemOpen
  \href@noop {} {}\bibinfo {note} {See {S}upplemental {M}aterial at [{URL}] for
  a detailed analysis of ${S}_{21}(\omega,x)$ within our model, the sample
  holder and for details on numerical simulations.}\BibitemShut {Stop}%
\bibitem [{\citenamefont {Khapaev}\ \emph {et~al.}(2002)\citenamefont
  {Khapaev}, \citenamefont {Kupriyanov}, \citenamefont {Goldobin},\ and\
  \citenamefont {Siegel}}]{khapaev2002current}%
  \BibitemOpen
  \bibfield  {author} {\bibinfo {author} {\bibfnamefont {M.~M.}\ \bibnamefont
  {Khapaev}}, \bibinfo {author} {\bibfnamefont {M.~Y.}\ \bibnamefont
  {Kupriyanov}}, \bibinfo {author} {\bibfnamefont {E.}~\bibnamefont
  {Goldobin}}, \ and\ \bibinfo {author} {\bibfnamefont {M.}~\bibnamefont
  {Siegel}},\ }\bibfield  {title} {\enquote {\bibinfo {title} {Current
  distribution simulation for superconducting multi-layered structures},}\
  }\href@noop {} {\bibfield  {journal} {\bibinfo  {journal} {Supercond. Sci.
  Technol.}\ }\textbf {\bibinfo {volume} {16}},\ \bibinfo {pages} {24}
  (\bibinfo {year} {2002})}\BibitemShut {NoStop}%
\bibitem [{\citenamefont {Pozar}(1998)}]{Poz98}%
  \BibitemOpen
  \bibfield  {author} {\bibinfo {author} {\bibfnamefont {D.~M.}\ \bibnamefont
  {Pozar}},\ }\href@noop {} {\emph {\bibinfo {title} {Microwave Engineering}}}\
  (\bibinfo  {publisher} {John Wiley \& Sons, Inc.},\ \bibinfo {year}
  {1998})\BibitemShut {NoStop}%
\bibitem [{\citenamefont {Ferdinand}\ \emph {et~al.}()\citenamefont
  {Ferdinand}, \citenamefont {Bothner}, \citenamefont {Koelle},\ and\
  \citenamefont {Kleiner}}]{ferdinand2017tunable}%
  \BibitemOpen
  \bibfield  {author} {\bibinfo {author} {\bibfnamefont {B.}~\bibnamefont
  {Ferdinand}}, \bibinfo {author} {\bibfnamefont {D.}~\bibnamefont {Bothner}},
  \bibinfo {author} {\bibfnamefont {D.}~\bibnamefont {Koelle}}, \ and\ \bibinfo
  {author} {\bibfnamefont {R.}~\bibnamefont {Kleiner}},\ }\bibfield  {title}
  {\enquote {\bibinfo {title} {Tunable superconducting {CPW} resonators using a
  lumped element {$\textrm{SrTiO}_3$} capacitor},}\ }\href@noop {} {\bibinfo
  {journal} {unpublished}\ }\BibitemShut {NoStop}%
\bibitem [{\citenamefont {Wenner}\ \emph {et~al.}(2014)\citenamefont {Wenner},
  \citenamefont {Yin}, \citenamefont {Chen}, \citenamefont {Barends},
  \citenamefont {Chiaro}, \citenamefont {Jeffrey}, \citenamefont {Kelly},
  \citenamefont {Megrant}, \citenamefont {Mutus}, \citenamefont {Neill},
  \citenamefont {O'Malley}, \citenamefont {Roushan}, \citenamefont {Sank},
  \citenamefont {Vainsencher}, \citenamefont {White}, \citenamefont {Korotkov},
  \citenamefont {Cleland},\ and\ \citenamefont
  {Martinis}}]{wenner2014catching}%
  \BibitemOpen
\bibfield  {journal} {  }\bibfield  {author} {\bibinfo {author} {\bibfnamefont
  {J.}~\bibnamefont {Wenner}}, \bibinfo {author} {\bibfnamefont
  {Y.}~\bibnamefont {Yin}}, \bibinfo {author} {\bibfnamefont {Y.}~\bibnamefont
  {Chen}}, \bibinfo {author} {\bibfnamefont {R.}~\bibnamefont {Barends}},
  \bibinfo {author} {\bibfnamefont {B.}~\bibnamefont {Chiaro}}, \bibinfo
  {author} {\bibfnamefont {E.}~\bibnamefont {Jeffrey}}, \bibinfo {author}
  {\bibfnamefont {J.}~\bibnamefont {Kelly}}, \bibinfo {author} {\bibfnamefont
  {A.}~\bibnamefont {Megrant}}, \bibinfo {author} {\bibfnamefont {J.~Y.}\
  \bibnamefont {Mutus}}, \bibinfo {author} {\bibfnamefont {C.}~\bibnamefont
  {Neill}}, \bibinfo {author} {\bibfnamefont {P.~J.~J.}\ \bibnamefont
  {O'Malley}}, \bibinfo {author} {\bibfnamefont {P.}~\bibnamefont {Roushan}},
  \bibinfo {author} {\bibfnamefont {D.}~\bibnamefont {Sank}}, \bibinfo {author}
  {\bibfnamefont {A.}~\bibnamefont {Vainsencher}}, \bibinfo {author}
  {\bibfnamefont {T.~C.}\ \bibnamefont {White}}, \bibinfo {author}
  {\bibfnamefont {A.~N.}\ \bibnamefont {Korotkov}}, \bibinfo {author}
  {\bibfnamefont {A.~N.}\ \bibnamefont {Cleland}}, \ and\ \bibinfo {author}
  {\bibfnamefont {J.~M.}\ \bibnamefont {Martinis}},\ }\bibfield  {title}
  {\enquote {\bibinfo {title} {Catching time-reversed microwave coherent state
  photons with 99.4\% absorption efficiency},}\ }\href@noop {} {\bibfield
  {journal} {\bibinfo  {journal} {Phys. Rev. Lett.}\ }\textbf {\bibinfo
  {volume} {112}},\ \bibinfo {pages} {210501} (\bibinfo {year}
  {2014})}\BibitemShut {NoStop}%
\bibitem [{\citenamefont {Bothner}\ \emph {et~al.}(2017)\citenamefont
  {Bothner}, \citenamefont {Wiedmaier}, \citenamefont {Ferdinand},
  \citenamefont {Kleiner},\ and\ \citenamefont
  {Koelle}}]{bothner2017improving}%
  \BibitemOpen
  \bibfield  {author} {\bibinfo {author} {\bibfnamefont {D.}~\bibnamefont
  {Bothner}}, \bibinfo {author} {\bibfnamefont {D.}~\bibnamefont {Wiedmaier}},
  \bibinfo {author} {\bibfnamefont {B.}~\bibnamefont {Ferdinand}}, \bibinfo
  {author} {\bibfnamefont {R.}~\bibnamefont {Kleiner}}, \ and\ \bibinfo
  {author} {\bibfnamefont {D.}~\bibnamefont {Koelle}},\ }\bibfield  {title}
  {\enquote {\bibinfo {title} {Improving superconducting resonators in magnetic
  fields by reduced field focussing and engineered flux screening},}\ }\href
  {\doibase 10.1103/PhysRevApplied.8.034025} {\bibfield  {journal} {\bibinfo
  {journal} {Phys. Rev. Applied}\ }\textbf {\bibinfo {volume} {8}},\ \bibinfo
  {pages} {034025} (\bibinfo {year} {2017})}\BibitemShut {NoStop}%
\end{thebibliography}%


\begin{thebibliography}{3}
\expandafter\ifx\csname natexlab\endcsname\relax\def\natexlab#1{#1}\fi
\expandafter\ifx\csname bibnamefont\endcsname\relax
  \def\bibnamefont#1{#1}\fi
\expandafter\ifx\csname bibfnamefont\endcsname\relax
  \def\bibfnamefont#1{#1}\fi
\expandafter\ifx\csname citenamefont\endcsname\relax
  \def\citenamefont#1{#1}\fi
\expandafter\ifx\csname url\endcsname\relax
  \def\url#1{\texttt{#1}}\fi
\expandafter\ifx\csname urlprefix\endcsname\relax\def\urlprefix{URL }\fi
\providecommand{\bibinfo}[2]{#2}
\providecommand{\eprint}[2][]{\url{#2}}

\bibitem[{\citenamefont{Pozar}(1998)}]{Poz98}
\bibinfo{author}{\bibfnamefont{D.~M.} \bibnamefont{Pozar}},
  \emph{\bibinfo{title}{Microwave Engineering}} (\bibinfo{publisher}{John Wiley
  \& Sons, Inc.}, \bibinfo{year}{1998}).

\bibitem[{\citenamefont{Deng et~al.}(2013)\citenamefont{Deng, Otto, and
  Lupascu}}]{deng2013}
\bibinfo{author}{\bibfnamefont{C.}~\bibnamefont{Deng}},
  \bibinfo{author}{\bibfnamefont{M.}~\bibnamefont{Otto}}, \bibnamefont{and}
  \bibinfo{author}{\bibfnamefont{A.}~\bibnamefont{Lupascu}},
  \bibinfo{journal}{J. Appl. Phys.} \textbf{\bibinfo{volume}{114}},
  \bibinfo{pages}{054504} (\bibinfo{year}{2013}).

\bibitem[{\citenamefont{Khapaev et~al.}(2002)\citenamefont{Khapaev, Kupriyanov,
  Goldobin, and Siegel}}]{khapaev2002current}
\bibinfo{author}{\bibfnamefont{M.~M.} \bibnamefont{Khapaev}},
  \bibinfo{author}{\bibfnamefont{M.~Y.} \bibnamefont{Kupriyanov}},
  \bibinfo{author}{\bibfnamefont{E.}~\bibnamefont{Goldobin}}, \bibnamefont{and}
  \bibinfo{author}{\bibfnamefont{M.}~\bibnamefont{Siegel}},
  \bibinfo{journal}{Supercond. Sci. Technol.} \textbf{\bibinfo{volume}{16}},
  \bibinfo{pages}{24} (\bibinfo{year}{2002}).

\end{thebibliography}

\end{document}


\renewcommand{\thefigure}{S\arabic{figure}}
\renewcommand{\theequation}{S\arabic{equation}}
\renewcommand{\thesection}{S\arabic{section}}


\title{Supplemental material for: Tunable superconducting two-chip lumped element resonator}

\author{B.~Ferdinand}
\email{benedikt-martin.ferdinand@uni-tuebingen.de}
\author{D.~Bothner}
\thanks{present adress: Kavli Institute of Nanoscience, Delft University of Technology, PO Box 5046, 2600 GA, Delft, The Netherlands}
\author{R.~Kleiner}%
\author{D.~Koelle}%

 \affiliation{%
Physikalisches Institut and Center for Quantum Science (CQ) in LISA$^{+}$,
Universit\"{a}t
T\"{u}bingen, Auf der Morgenstelle 14, D-72076 T\"{u}bingen, Germany
}%
\date{\today}

%
\pacs{05.45.-a,  
      05.40.-a,   
      05.60.Cd,  
      74.50.+r  
      } %

\maketitle

\section{Model}
\label{sec:details}
\begin{figure}[ht]
			\includegraphics[width=0.6\textwidth]{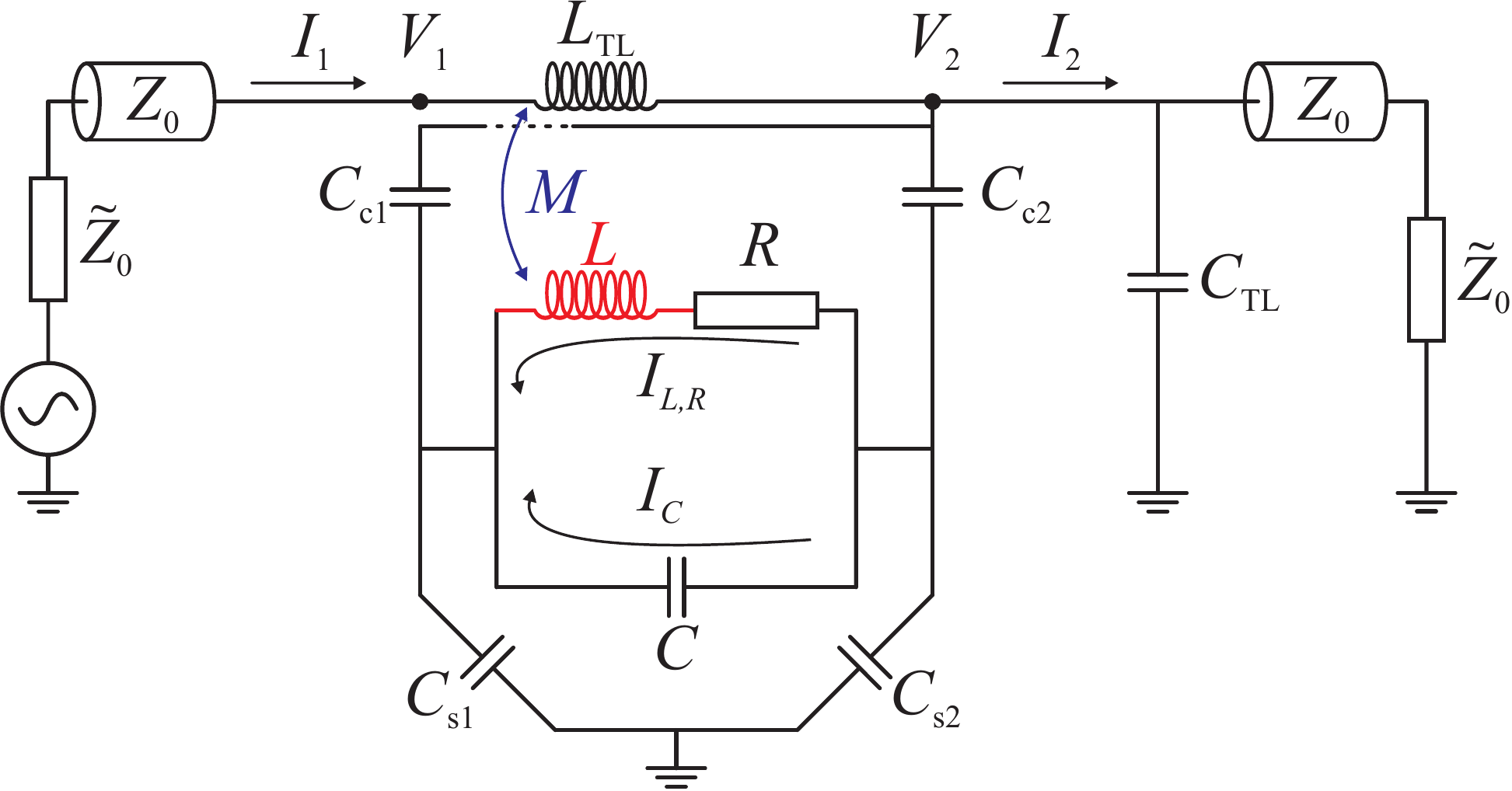} 
	\caption{Circuit model used to describe the two-chip resonator.}
	\label{fig:Scheme_Suppl}
\end{figure}
Solving Kirchhoff equations (without the capacitor $C_\textrm{TL}$ and the impedance $\tilde{Z}_0$) for the circuit shown in Fig.~\ref{fig:Scheme_Suppl} and assuming $C_\textrm{c1}=C_\textrm{c2}$ and $C_\textrm{s1}=C_\textrm{s2}$ leads to
\begin{equation} \label{eq:kirchhoff}
\begin{split}
V_2&=\frac{1}{i\omega C_c}\bigg\{\frac{V_1-V_2+i\omega MI_{L,R}}{i\omega L_\textrm{TL}}-I_2-\frac{1}{2}\bigg\{\frac{V_1-V_2+i\omega MI_{L,R}}{i\omega L_\textrm{TL}}-I_2 \\
&+\bigg\{(i\omega L+R)I_{L,R}-i\omega M\frac{V_1-V_2+i\omega MI_{L,R}}{i\omega L_\textrm{TL}}\bigg\}\cdot i\omega C_\textrm{c}\bigg\}\bigg\}+(i\omega L+R)I_{L,R}-i\omega M I_1 \\
&+\frac{1}{i\omega C_\textrm{s}}\bigg\{\frac{1}{2}\frac{V_1-V_2+i\omega MI_{L,R}}{i\omega L_\textrm{TL}}-I_2+\bigg\{(i\omega L+R)I_{L,R}-i\omega M\frac{V_1-V_2+i\omega MI_{L,R}}{i\omega L_\textrm{TL}}\bigg\}\cdot i\omega C_\textrm{c}\bigg\}\bigg\} \\
&+I_{L,R}+\bigg\{(i\omega L+R)I_{L,R}-i\omega M\frac{V_1-V_2+i\omega MI_{L,R}}{i\omega L_\textrm{TL}}\bigg\}\cdot i\omega C, \\ \\
I_1&=\frac{V_1-V_2+i\omega MI_{L,R}}{i\omega L_\textrm{TL}}, \\ \\
I_{L,R}&=\frac{i\omega M[(V_2-V_1)\cdot(2C+C_\textrm{c}+C_\textrm{s})]}{-2L_\textrm{TL}-i\omega L_\textrm{TL}R(2C+C_\textrm{c}+C_\textrm{s})+\omega^2(LL_\textrm{TL}-M^2)\cdot(2C+C_\textrm{c}+C_\textrm{s})}.
\end{split}
\end{equation}
Using these expressions (\ref{eq:kirchhoff}), one can evaluate the matrix elements of the $ABCD$ matrix $\hat{S}_\textrm{LE}$ of the lumped element resonator via
\begin{equation} \label{eq:ABCD}
	\begin{pmatrix}
	V_1 \\
	I_1 \\
	\end{pmatrix}=
	\begin{pmatrix}
	A & B \\
	C & D \\
	\end{pmatrix}
	\begin{pmatrix}
	V_2 \\
	I_2 \\
	\end{pmatrix}=\hat{S}_\textrm{LE}
	\begin{pmatrix}
	V_2 \\
	I_2 \\
	\end{pmatrix}.
\end{equation}
Multiplying this matrix with the $ABCD$ matrix 
\begin{equation}
\hat{S}_{C\textrm{TL}}=
	\begin{pmatrix}
	1 & 0 \\
	i\omega C_\textrm{TL} & 1 \\
	\end{pmatrix}
\end{equation}
of the capacitor $C_\textrm{TL}$, one gets the total $ABCD$ matrix of the circuit, and hence, the transmission $S$-parameter $S_{21}(\omega)$ by using \cite{Poz98}
\begin{equation} \label{eq:Sparam}
S_{21}=\frac{2}{A+B/Z_0+CZ_0+D}.
\end{equation}

The uncoupled resonance frequency of the device is $\omega_\textrm{u}=(\sqrt{LC})^{-1}=2\pi\cdot 5.98\,\textrm{GHz}$ ($L=1129\,\textrm{pH}$, $C=626\,\textrm{fF}$). However, due to the capacitive and inductive coupling and the presence of the superconducting surfaces of the TL, the frequency is shifted significantly. In order to get analytical expressions for the coupled resonance frequency and the quality factors, approximations are made in the following. By assuming a high internal quality factor $Q_i>10^3$, and consequently a small resistance $R$ of the resonator, one can expect little influence of the resistance on the resonance frequency. Thus, we calculate the resonance frequency $\omega_r$ of the resonator using $R=0$, which leads to the form
\begin{equation} \label{eq:Poly}
S_{21,R=0}=\frac{a_0+a_2\omega^2}{\sum_{j=1}^4 b_j\omega^j}\,\, ,
\end{equation}
where $a_0,a_2$ are real values and the $b_j$ are complex values, which depend on the circuit parameters. We then describe our reduced circuit with dimensionless parameters,
\begin{equation} \label{eq:dimless}
\mu=\frac{M}{L} ,\ \ \lambda=\frac{L_\textrm{TL}}{L} ,\ \ \gamma_c=\frac{C_\textrm{c}}{C} ,\ \ \gamma_s=\frac{C_\textrm{s}}{C} ,\ \ \gamma_\textrm{TL}=\frac{C_\textrm{TL}}{C} ,\ \ \zeta=\frac{\sqrt{L/C}}{Z_0}\,\, ,
\end{equation}
and furthermore define $\gamma=\gamma_c+\gamma_s$. By setting the numerator of the transmission function to zero, we obtain two solutions for $\omega_\textrm{r}$, one of which is the negative of the other. The solution for $\omega_\textrm{r}>0$ is shifted to smaller values, compared to the uncoupled resonance frequency $\omega_\textrm{u}=1/\sqrt{LC}$. We find the resonance frequency in absence of internal losses to be given by
\begin{equation} \label{eq:fres}
\omega_\textrm{r}=\frac{\omega_\textrm{u}}{\sqrt{1+\gamma/2}}
\end{equation}
According to Eq.~(\ref{eq:fres}), the resonance frequency depends on $L$, $C$ and the sum of the coupling capacitances $C_\textrm{c}$ and $C_\textrm{s}$. As shown in the main manuscript, when $x_\textrm{LE}$ is varied, the sum of them remains almost constant. Thus, the frequency changes are barely induced by these capacitances. The resonance frequency is thus shifted by the varying inductance $L$ for different positions $x_\textrm{LE}$.
 
For the implementation of internal losses into the model, we then transform $i\omega L\rightarrow i\omega L+R$, or equivalently, the inductance $L$ to
\begin{equation} \label{eq:Ltrans}
L\rightarrow L+\frac{R}{i\omega}=L(1-iq),
\end{equation}
where we have used the definition of the internal quality factor $Q_\textrm{i}$ of the $LCR$ circuit presented in the main manuscript
\begin{equation} \label{eq:Qfac}
q=\frac{1}{Q_\textrm{i}}=\frac{R}{\omega_\textrm{r}L}.
\end{equation}
In Eq.~(\ref{eq:Qfac}) we have assumed that the system is considered only close to resonance ($\omega\approx\omega_\textrm{r}$).

We expand our derived expression for $S_{21}$ around $\omega_\textrm{r}$ up to the linear order in $\Delta \omega=\omega-\omega_\textrm{r}$ for both the numerator and the denominator and receive as an approximation, analogous to \cite{deng2013},
\begin{equation} \label{eq:S21}
S_{21}(\omega)=Ae^{i\phi}\frac{1+2iQ_\textrm{i}\frac{\Delta \omega}{\omega_\textrm{r}}}{1+\frac{Q_\textrm{i}}{Q_\textrm{e}}+i\frac{Q_\textrm{i}}{Q_\textrm{a}}+2iQ_\textrm{i}\frac{\Delta \omega}{\omega_\textrm{r}}}.
\end{equation}

\begin{figure}[ht]
			\includegraphics[width=0.75\textwidth]{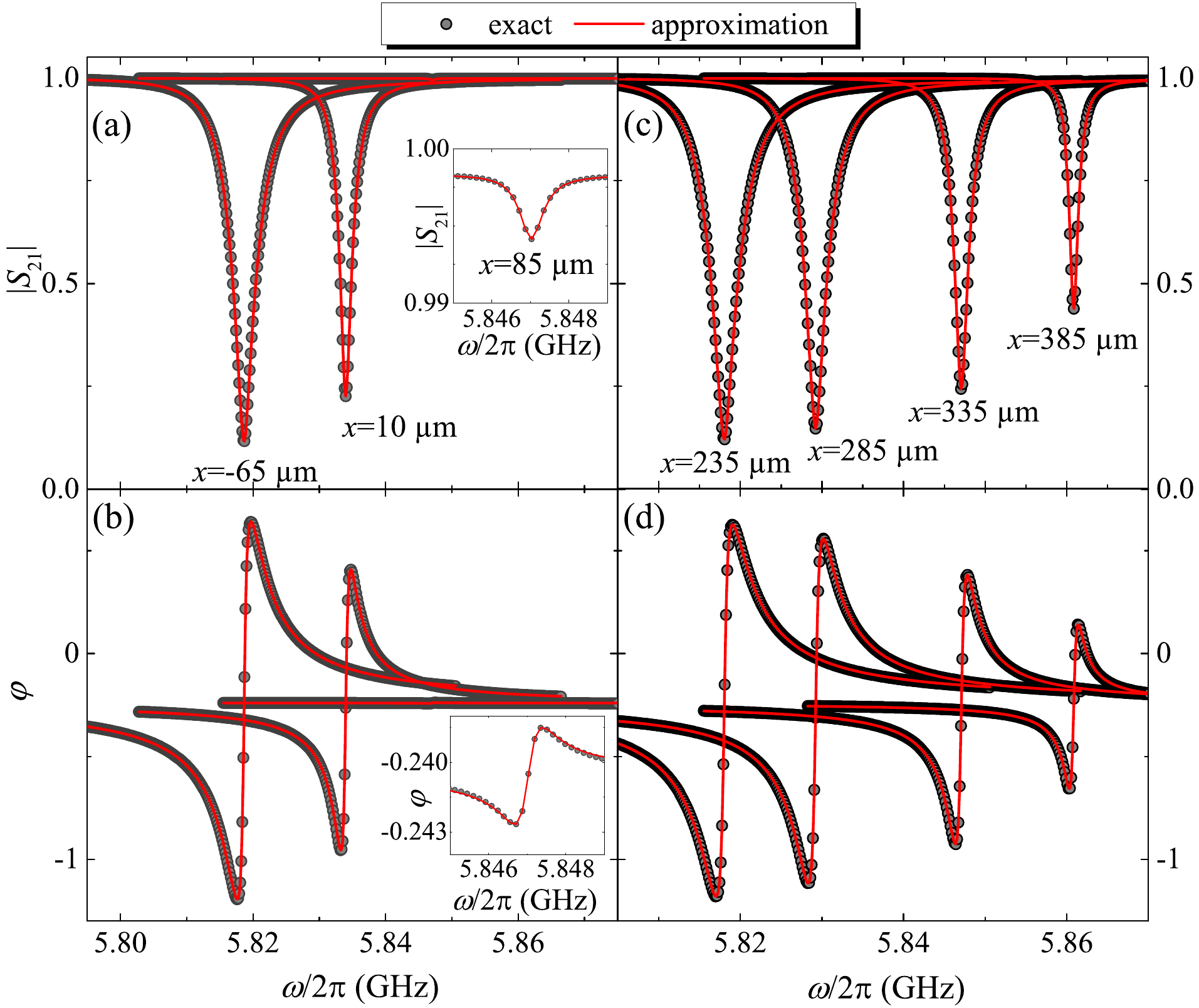} 
	\caption{Comparison between the exact model (grey circles) and Eq.~(\ref{eq:S21}) (red lines). (a), (c) show the amplitude, (b) and (d) the phase of the forward scattering parameter $S_{21}(\omega)$ for various positions $x_\textrm{LE}$.}
	\label{fig:Model_2}
\end{figure}

In order to demonstrate the validity of this approximation, for selected values of the position $x_\textrm{LE}$, both the exact (grey circles) and approximated (red lines) amplitude and phase of the forward scattering parameter $S_{21}(\omega)$ are shown in Fig.~\ref{fig:Model_2} as a function of frequency $\omega$.
We find excellent agreement for all experimentally relevant regimes, specified in the main manuscript. Thus, fits to measurement data using Eq.~(\ref{eq:S21}) is valid for all of these regimes.

Performing a multivariable Taylor expansion for both, the numerator and the denominator of Eq.~(\ref{eq:S21}) separately in orders of $\mu$, $\gamma_c$ and $\gamma_s$, one approximately gets
\begin{equation}
Q_\textrm{e}\approx\frac{\gamma_\textrm{TL}^2+4\zeta^2-2\gamma_\textrm{TL}\lambda\zeta^2+\zeta^4\lambda^2+\zeta^2\gamma_\textrm{TL}^2\lambda^2}{(2\zeta^2+\gamma_\textrm{TL}^2)\zeta\mu^2}
\end{equation}
and
\begin{equation}
Q_\textrm{a}\approx\frac{\gamma_\textrm{TL}^2+4\zeta^2-2\gamma_\textrm{TL}\lambda\zeta^2+\zeta^4\lambda^2+\zeta^2\gamma_\textrm{TL}^2\lambda^2}{(\gamma_\textrm{TL}-\gamma_\textrm{TL}^2\lambda-\zeta^2\lambda)\zeta^2\mu^2}
\end{equation}
In our model, we neglect the impedance $\tilde{Z}_0=50\,\Omega$ and calculate the transmission function using the impedance $Z_0$ in Eq.~\ref{eq:Sparam}. Thus, the calculated transmission function and the corresponding external quality factors can be seen as measured next to the resonator structure. Using $\tilde{Z}_0$ instead of $Z_0$ reduces the calculated external quality factors $Q_\textrm{e}$ to $\approx85\%$ of the values shown in the main manuscript. 
\clearpage

\section{Measurement setup}
\begin{figure}[ht]
			\includegraphics[width=0.45\textwidth]{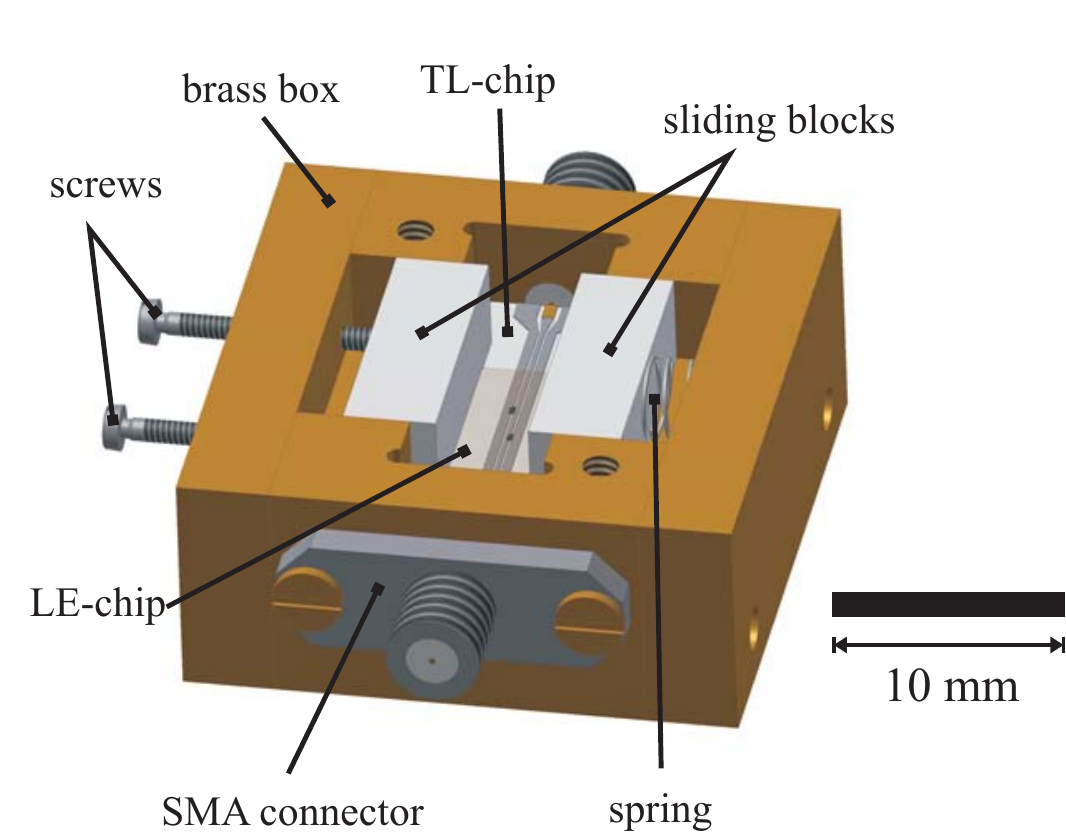} 
	\caption{Sample holder used for varying the position $x_\textrm{LE}$ (for details see text).}
	\label{fig:SampleHolder}
\end{figure}
In the experiment at $T=4.2\,\textrm{K}$, we change the position of the resonator with the sample holder shown in Fig.~\ref{fig:SampleHolder}. It consists of a brass box, in which the TL with finite ground planes is mounted. The TL is connected to the input and output circuit via SMA connectors. On top of the TL-chip, the LE-chip is clamped between two sliding blocks, which have small millings such that the LE-chip is tightly attached to the TL-chip. A lid closes the sample holder (not shown). The position can be changed by rotating the screws, whereas the spring takes care for a small pressure on the LE-chip, which accordingly can be moved in both directions. The chip stacking process is done under an optical microscope and the accuracy in the determination of the relative distance $x_\textrm{LE}$ is better than $\pm5\,\mathrm{\mu m}$.

\section{Details of the simulation}
\begin{figure}[ht]
			\includegraphics[width=0.75\textwidth]{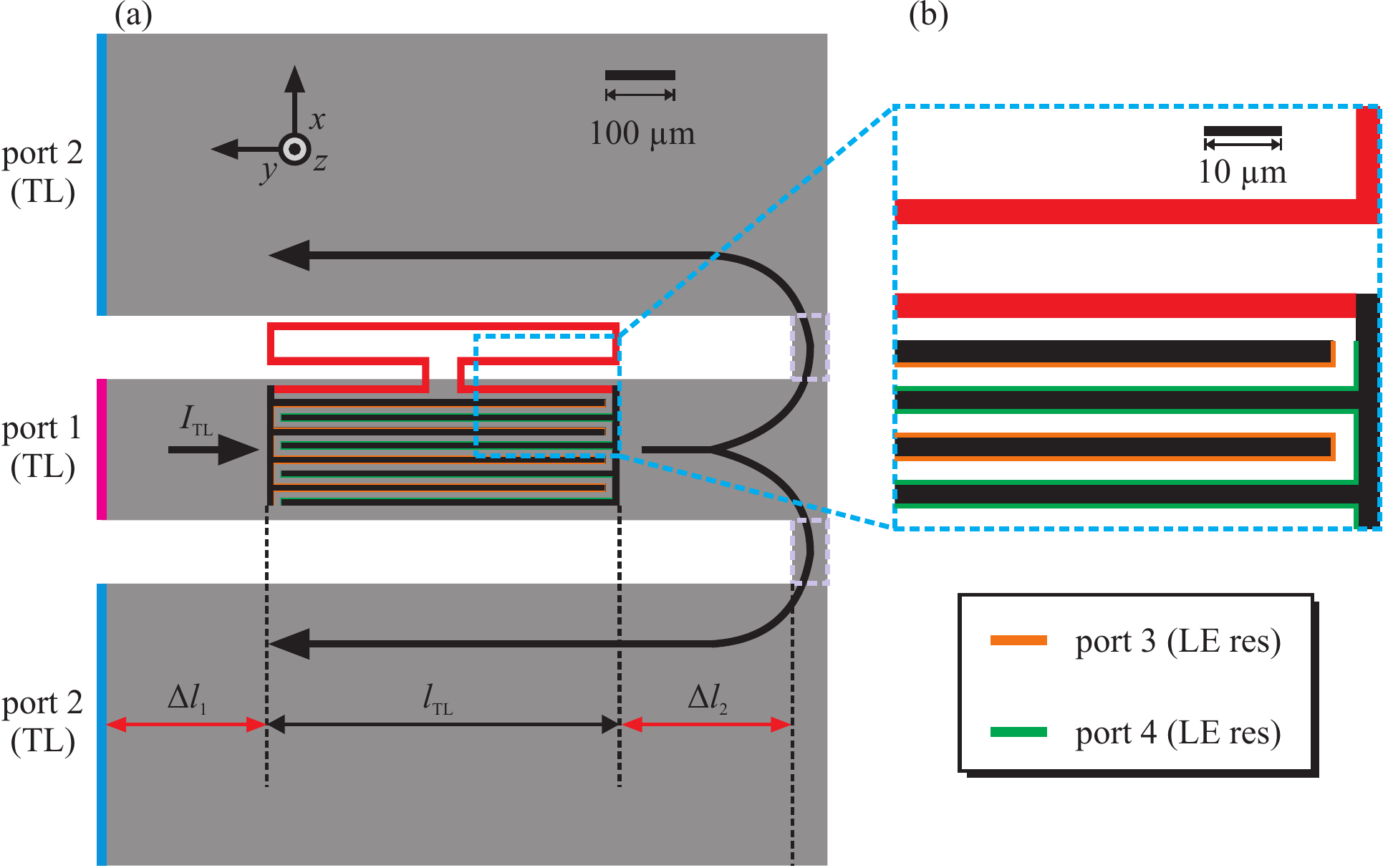} 
	\caption{Schematic of the boundary conditions used to calculate the inductance matrix numerically. The whole geometry is shown in (a), consisting of the TL part (grey) and the LE resonator with the inductive loop (red) and the IDC (black). For the simulation of the inductance matrix, the TL current $I_\textrm{TL}$ (direction indicated by the black arrows) flow from port 1 to port 2, whereas the currents of the LE resonator go from port 3 to port 4. The grey dashed rectangle frames inductive shorts from the center conductor of the TL to the ground planes. Panel (b) shows an enlarged picture of the IDC in order to make the choice of boundary conditions for the currents of the LE resonator more clear.}
	\label{fig:DetailsSimu}
\end{figure}
For the simulation of the inductive part of the circuit we use the finite element software 3D-MLSI \cite{khapaev2002current}. Based on the London and the Maxwell equations, this software enables the simulation of circuits in 2.5 dimensions, meaning that the boundary conditions are applied at the edges of a two-dimensional structure ($x$- and $y$-direction), and in the remaining $z$-direction, effects of the generated magnetic fields and the corresponding inductances and mutual inductances are taken into account using the Biot-Savart law. A schematic of the circuit is drawn in Fig.~\ref{fig:DetailsSimu}(a), with an enlarged version in (b) focusing on the LE resonator. The figure illustrates the boundary conditions for current paths used to simulate the inductance matrix. The currents in the TL are defined to flow from the left hand side of the center conductor (port 1 (TL)), across the inductive shorts (grey dashed frames) to the left hand side of the ground planes (port 2 (TL)). Of course, these shorts are not present in the real TL structure, however, including them this way for the simulations automatically maps both, the inductance of the ground planes and the mutual inductance between center conductor and ground plane, onto the TL inductance, as typical for the description in the framework of TL theory \cite{Poz98}. The distance $\Delta l_2=250\,\mathrm{\mu m}$ between the right end of the LE resonator and the shorts is chosen as a compromise, such that, on the one hand, no significant amount of flux generated by the currents along these shorts is threading the inductive loop of the LE resonator, and on the other hand, the mesh is small enough to be able to simulate the structure. In addition, the inductance of the shorts does not change the TL inductance by much. We find that the presence of the LE resonator does not affect the inductance of the TL. Hence, the simulated value of the TL can easily be adjusted to the correct length of the LE resonator along the TL in order to get $L_\textrm{TL}$ of the main manuscript. 

The capacitance matrix is calculated using COMSOL Multiphysics, which offers the possibility of a full 3D simulation. For the simulation, only the IDC part of the LE resonator (black in Fig.~\ref{fig:DetailsSimu}) is taken into account and the inductive loop structure is absent (red in Fig.~\ref{fig:DetailsSimu}). Thus, the two sides of the capacitor are galvanically separated for the simulations. Furthermore, the inductive shorts of the TL (grey dashed frames) are not present in this simulation. Similar as for the TL inductance $L_\textrm{TL}$, also the matrix entry $C_\textrm{TL}'$ corresponding to the TL capacitance $C_\textrm{TL}$ has to be adjusted to the length of the resonator after the simulation. This is done by calculating the capacitance per unit length $C_l$ of the TL cross-section with the second substrate on top, but without any LE resonator structure. Afterwards the capacitance $C_l\cdot(\Delta l_1+\Delta l_2)$ is subtracted from $C_\textrm{TL}'$ on both sides of the TL to get $C_\textrm{TL}=C_\textrm{TL}'-C_l\cdot(\Delta l_1+\Delta l_2)$ (cf. Fig.~\ref{fig:DetailsSimu} for definition of $\Delta l_1$ and $\Delta l_2$). 

\bibliography{Ferdinand_LE-Model}